\documentclass{article}

\usepackage{arxiv}

\usepackage[utf8]{inputenc} % allow utf-8 input
\usepackage[T1]{fontenc}    % use 8-bit T1 fonts
\usepackage{hyperref}       % hyperlinks
\usepackage{url}            % simple URL typesetting
\usepackage{booktabs}       % professional-quality tables
\usepackage{amsfonts}       % blackboard math symbols
\usepackage{nicefrac}       % compact symbols for 1/2, etc.
\usepackage{microtype}      % microtypography
\usepackage{lipsum}		% Can be removed after putting your text content
\usepackage{graphicx}
\usepackage{natbib}
\usepackage{doi}

\usepackage{svg}
\usepackage{changepage}
\usepackage{csquotes}
\usepackage{enumitem}
\usepackage{multicol}
\usepackage{amsfonts}
\usepackage{listings}
\usepackage{caption}
\usepackage{subcaption}
\usepackage{titlesec}
\usepackage{appendix}
\usepackage{xargs}       

\usepackage{bookmark}
\usepackage{soul}
\usepackage{amsthm}
\usepackage{amsmath}
\usepackage{amsfonts}
\usepackage{listings}
\usepackage{float}
\usepackage{caption}
\usepackage{wrapfig}
\usepackage{subcaption}
\usepackage{titlesec}
\usepackage{here}
\usepackage{colortbl}
\usepackage{array}
\newcolumntype{L}{>{\centering\arraybackslash}p{0.15\linewidth}}

\newcommand{\dint}{\delta_{int}}
\newcommand{\dext}{\delta_{ext}}

\date{} 					% Or removing it

\author{
  \hspace{1mm} Daniel J. Foguelman \\
CONICET-Universidad de Buenos Aires\\
  Instituto de Investigación en \\
  Ciencias de la Computación (ICC)\\
  Universidad de Buenos Aires\\
  Facultad de Ciencias Exactas y Naturales\\
  Departamento de Computación.\\
  Buenos Aires, Argentina.\\
  \texttt{dfoguelman@dc.uba.ar} \\
  \And
   \hspace{1mm} Esteban Lanzarotti \\
  CONICET-Universidad de Buenos Aires\\
  Instituto de Investigación en \\
  Ciencias de la Computación (ICC)\\
  Universidad de Buenos Aires\\
  Facultad de Ciencias Exactas y Naturales\\
  Departamento de Computación.\\
  Buenos Aires, Argentina.\\
  \texttt{elanzarotti@dc.uba.ar} \\
   \And
   \hspace{1mm} Emanuel Ferreyra \\
   CONICET-Universidad de Buenos Aires \\
   Universidad de Buenos Aires \\
   Facultad de Ciencias Exactas y Naturales\\
   Instituto de C{\'a}lculo (IC).\\
   Buenos Aires, Argentina.\\
   \texttt{emanuel.ferreyra@ic.fcen.uba.ar} \\
     \And 
  \hspace{1mm} Rodrigo Castro \\
  CONICET-Universidad de Buenos Aires\\
  Instituto de Investigación en \\
  Ciencias de la Computación (ICC)\\
  Universidad de Buenos Aires\\
  Facultad de Ciencias Exactas y Naturales\\
  Departamento de Computación.\\
  Buenos Aires, Argentina.\\
   \texttt{rcastro@dc.uba.ar} \\
}

\title{Simulation of emergence in artificial societies:\\ a practical model-based approach\\ with the EB-DEVS formalism}

\usepackage{hyperref}
\newcommand*{\fullref}[1]{\hyperref[{#1}]{\autoref{#1}}} % One single link

\definecolor{codegreen}{rgb}{0,0.6,0}
\definecolor{codegray}{rgb}{0.5,0.5,0.5}
\definecolor{codepurple}{rgb}{0.58,0,0.82}
\definecolor{backcolour}{rgb}{0.95,0.95,0.92}
\definecolor{babyblueeyes}{rgb}{0.63, 0.79, 0.95}
\definecolor{bazaar}{rgb}{0.6, 0.47, 0.48}

\lstdefinestyle{base}{
    basicstyle=\scriptsize\ttfamily,
    breakautoindent=true,
    language=Python,
    breakatwhitespace=false,
    breaklines=true,
    captionpos=b,
    commentstyle=\itshape\color{codegreen},
    firstnumber=1,
    keepspaces=true,
    keywordstyle=\bfseries,
    linewidth=\columnwidth,
    breaklines=true,
    numberfirstline=false,
    numbers=left,
    numbersep=5pt,
    numberstyle=\tiny\color{codegray},
    showspaces=false,
    showstringspaces=false,
    showtabs=false,
    stepnumber=1,
    stringstyle=\color{codepurple},
    tabsize=2,
    xleftmargin=2em,
     escapeinside={||},
    mathescape=true,
    moredelim=**[is][\color{red}]{@}{@},
    moredelim=*[is][\color{blue}]{/*}{/*}, 
    columns=fullflexible,
    escapeinside={||},
    mathescape=true,
    morekeywords={AgentState, AgentAtomicExternalTransition,AgentAtomicInternalTransition,AgentAtomicOutputFunction,AtomicLambdaFunction,CoupledGlobalTransition,CoupledState,CellAtomicTimeAdvance,CellAtomicOutputFunction,CellAtomicInternalTransition,CellAtomicExternalTransition,AgentAtomicTimeAdvance,NodeAtomicInternalTransition,CoupledModelTransition,Gini, CellState}
}
\begin{document}
\maketitle 
%%%%%%%%%%%%%%%%%%%%%%%%%%%%%%%%%%%%%%%%%%%%%%
% Abstract and keywords
\begin{abstract}
Modelling and simulation of complex systems is key to exploring and understanding social processes, benefiting from formal mechanisms to derive global-level properties from local-level interactions.
In this paper we extend the body of knowledge on formal methods in complex systems by applying EB-DEVS, a novel formalism tailored for the modelling, simulation and live identification of emergent properties.
We guide the reader through the implementation of different classical models for varied social systems to introduce good modelling practices and showcase the advantages and limitations of modelling emergence with EB-DEVS, in particular through its live emergence detection capability.
This work provides case study-driven evidence for the neatness and compactness of the approach to modelling communication structures that can be explicit or implicit, static or dynamic, with or without multilevel interactions, and with weak or strong emergent behaviour.
Throughout examples we show that EB-DEVS permits conceptualising the analysed societies by incorporating emergent behaviour when required, namely by integrating as a macro-level aggregate the Gini index in the Sugarscape model, Fads and Fashion in the Dissemination of Culture model, size-biased degree distribution in a Preferential Attachment model, happiness index in the Segregation model and quarantines in the SIR epidemic model.
We also show in practice that the formalism and its implementation are backward compatible with Classic DEVS, allowing the incremental extension of previously developed models with  new features provided by EB-DEVS.
In each example we discuss the role of communication structures in the development of multilevel simulation models, and illustrate how micro-macro feedback loops enable the modelling of macro-level properties. Our results stress the relevance of multilevel features to support a robust approach in the modelling and simulation of complex systems.
\end{abstract}


% \begin{keywords}
% {}
% \end{keywords}

%%%%%%%%%%%%%%%%%%%%%%%%%%%%%%%%%%%%%%%%%%%%%%
% Start of  paragraph numbering. Please leave this untouched
% \parano{}

\section{Introduction}

Agent Based Modelling (ABM) has proven a very successful simulation-driven tool to explain phenomena, discover new questions and challenge prevailing theories. 
In \textit{Why model?} \citep{Epstein2008WhyModel} the author proposes a comprehensive guide in a broad spectrum of good reasons to model. The social sciences have found computer simulation to be a valuable  technique for experimenting with social models and hypothetical scenarios. Noteworthy models in the literature cover multiple disciplines such as political science \citep{Axelrod1997,Schelling1971}, economics \citep{Tesfatsion2002Agent-basedUp.}, sociology \citep{Macy2002FromModeling} and epidemiology \citep{Kermack1927AEpidemics}, just to name a few.

According to \citet{Edmonds2005ComputationalExperiment} \textit{``[models] play a supporting role to the main argument. In many cases, if a paper that uses such a model were rewritten without it, the conclusions would not greatly differ, but rather the formal model is a concrete expression and demonstration of the processes being discussed''} . 

Research in artificial societies benefits from formal modelling as it supports new theories in a concise, correct and unambiguous fashion. According to \citet{Sawyer2005SocialSystems} complex systems' theory is key to understand complex dynamical systems such as societies and emergence is a key element that allows for multiple levels of analysis. Consequently, integrating higher and lower levels of abstraction with powerful tools may help the modeller to gain deeper understanding of the system under study. 

In this paper we extend the body of knowledge on formal methods in complex systems by applying a novel formalism tailored for the modelling, simulation and live identification of emergent properties.

% This is vital when modelling complex systems. Natural language cannot provide enough precision to explain the thought process behind the model, and ad-hoc modelling methods hinders expressiveness due to implementation complexities. Similar to natural language for modelling, ad-hoc methods are not enough either.

% We can define complex systems as interacting dynamic components that may produce novel system-level properties that cannot be directly explained by analysing components in isolation~\cite{bar2002general}. Hence, emergence and complex systems are bonded concepts. John Holland, has used emergent properties as a \textit{sufficient property} to characterize a system as complex~\cite{Holland2006}.

% In the Springer Complexity program, a complex system is defined as:

% \begin{displayquote}
% \textit{``[\ldots] systems that comprise many interacting parts with the ability to generate a new quality of macroscopic collective behaviour the manifestations of which are the spontaneous formation of distinctive temporal, spatial or functional structures.''} \cite{Springer2020SpringerProgram}
% \end{displayquote}

Emergent patterns, structures and properties need to be conceptualised. Authors from the social sciences have acknowledged that top-down approaches are inefficient due to the extensive knowledge needed for its formalisation, while bottom-up modelling provides better means to understand social systems. Agent based modelling, helps to understand global properties as emergent properties of actors interactions. In Macy's words: % \begin{displayquote}
\textit{``we may be able to understand these dynamics much better by trying to model them, not at the global level but instead as emergent properties of local interaction among adaptive agents who influence one another in response to the influence they receive.''} \citet[p.~144]{Macy2002FromModeling}
% \end{displayquote} 

In this work we introduce good modelling practices for complex systems with EB-DEVS \citep{Foguelman_2021}, a formalism designed to deal explicitly with emergent properties. We guide the reader through the implementation of classical models in social sciences based on this formalism. Our goal is to showcase the advantages and limitations of modelling emergence with EB-DEVS, in particular through its live detection capabilities.

Formal Modelling and Simulation (M\&S) approaches facilitate the interpretation and reuse of simulation models by means of clear unambiguous semantics. The Discrete Event System Specification (DEVS) is a modelling formalism for discrete-event systems capable of representing exactly any discrete system, and of approximating continuous systems with any desired accuracy. DEVS also makes emphasis on modular and hierarchical composition of (possibly heterogeneous) subsystems. 
In addition, its clear separation of concerns between model definition and model execution will allow us to focus strictly on the modelling aspects first, leaving the intricacies of executing a discrete event simulation model as a separate, though very important second stage. 

DEVS \citep{Zeigler2018TheoryFoundations} has been extensively used for the modelling of complex systems due its capabilities to represent a plethora of formalisms, both in the discrete and continuous realms, and also in deterministic and stochastic domains. The relevance of emergent behaviour under the perspective of DEVS M\&S, is exposed with great detail in \citet{Mittal2013} where the author elaborates on how DEVS extensions can be used to tackle emergence, stigmergy and complex adaptive systems modelling. The M\&S community has worked extensively in the modelling, ex-post identification (after a simulation is completed) and validation of emergent properties. Nevertheless, few of these efforts were effective to provide a formal, system-theoretic M\&S approach to deal with emergence in the live system (a review can be found in \citet{Szabo2017}).

In a recent work \citep{Foguelman_2021} we successfully extended DEVS for the modelling of strong emergence in complex adaptive systems. This new formalism allows for the modelling of complex interactions between multiple levels. Atomic models (the DEVS representation for agents and actors) access information provided horizontally (between same level models) or hierarchically (from higher level models). The extension implements upward causation and downward information mechanisms allowing the live-detection and implementation of emergent properties.
The formalism and its implementation are backward compatible with existing DEVS simulators allowing the extension of already implemented models with the new features provided by EB-DEVS. 

We develop a series of classical models that point out several features that we consider fundamental to the goal of establishing best practices and guidelines in the modelling of complex systems with emergent properties. The cherry picked models presented in this work explore the realms of strong and weak emergence, adaptive agents with explicit and implicit communication using static and dynamic topologies. We show an overview of the implemented features and models in \autoref{tbl:models_features}.

The structural features that are present in our models, can be divided into five categories depending on their communication type, if explicit or implicit; if the model topology changes through time, giving the model a dynamic or static communication structure; what multilevel channels are used, from the bottom-up or from the top-down; the type of emergence integrated in the model, if weak or strong; and by the emergence identification process in the model, whether a live identification is done in simulation time or if it is done after the simulation is over. 

The rest of the manuscript is structured as follows: in \autoref{sec:modelling_complex_systems} we introduce the reader with two M\&S formalisms: DEVS and its extension EB-DEVS. To do this we show two simple examples that helps to understand how these formalisms work. Furthermore, we present and describe how several properties of complex systems are implemented and manipulated  through a series of examples. Then in \autoref{sec:models} we present the models together with their implementations, experimental results, and discussions on their attributes as complex systems. We close the work with a discussion regarding the contributions of this new formalism in \autoref{sec:discussion} and \autoref{sec:conclusions}.

\section{Modelling complex systems with emergent properties}
\label{sec:modelling_complex_systems}

In this section we provide a brief introduction to DEVS, a formalism for the M\&S of general systems. We then review the EB-DEVS extension, providing a simple example to showcase the key primitives used to model an abstract system. Finally, we provide a list of features in complex systems that can be modelled with EB-DEVS, and that will be present throughout the examples in \autoref{sec:models}.

% TODO: A brief discussion regarding how DEVS has been used to model complex systems
\subsection{DEVS}
A DEVS model processes an input event trajectory and, according to that trajectory and/or its own internal state, can undergo state changes and produce an output event trajectory.

Formally, a DEVS \emph{atomic} model $M_{DEVS}$ is defined by the following structure:
\begin{eqnarray}
  M_{DEVS}= \: \langle X,Y,S,\dint,\dext,\lambda,ta \rangle
\end{eqnarray}
where  $X$ is the set of input event values, $Y$ is the set of output event values and $S$ is the set of state values.  $\dint$, $\dext$, $\lambda$ and $ta$ are the functions that  define the model dynamics.

Each possible state $s$ ($s\in S$) has an associated \emph{time advance} calculated by the \emph{time advance function} $ta(s)$ ($ta(s):S \rightarrow \Re^+_0$) that determines how long the model will remain in state $s$ in the absence of external input events.
If an Atomic model is in state $s_1$ at time $t_1$, after $ta(s_1)$ units of time (i.e.\ at time $ta(s_1)+t_1$) the system
performs an \emph{internal transition}, resulting in a new state $s_2$.
The new state is calculated as $s_2=\dint(s_1)$, where $\dint$ ($\dint:S\rightarrow S$) is called \emph{internal transition function}. Before this state transition from $s_1$ to $s_2$ an output event is produced with value $y_1=\lambda (s_1)$, where $\lambda$ $(\lambda:S \rightarrow Y)$ is called \emph{output function}.

When an input event arrives, the external state transition function $\dext$ ($\dext:S\times
\Re^+_0 \times X \rightarrow S$) is invoked. 
The  new state value depends not only on the input event value but also on the previous state value and the elapsed time since the last transition.

In \autoref{example1} we show a visually intuitive example for a typical evolution of events and evaluations of functions in an Atomic DEVS model.

DEVS models can be coupled modularly and hierarchically allowing the definition of multilevel systems and the reuse of previously defined modules. Formally, a DEVS \emph{coupled} model $CN_{DEVS}$ (also a "Coupled Network" of DEVS models) is defined by the following structure:
\begin{equation}
  CN_{DEVS}= \:\langle X_{self},Y_{self},D,\{M_i\},\{I_i\},\{Z_{i,j}\}, Select \rangle
\end{equation} 

where:
 $X_{self}$ and $Y_{self}$ are the sets of input and output values of the coupled model. $D$ is the set of component references, so that for each $d \in   D$, $M_d$ is a DEVS model. For each $d\in D\cup \{self\}$, $I_i$ is the set of Influencer models on subsystem $d$. For each $i\in I_i$, $Z_{i,j}$ is the translation function, while  $Select:2^D\rightarrow D$ is a tie--breaking function for simultaneous events.

DEVS models are closed under coupling, i.e., the coupling of DEVS
models defines an equivalent atomic DEVS model \citet{Zeigler2018TheoryFoundations}.  

Classic DEVS can represent a wide variety of formalisms \citep{vangheluweDEVSMUlti} from the discrete and continuous realms, to the deterministic and stochastic domains. Since its introduction in \citet{zeigler1976theory} several new practical needs arose from the challenges in the practice of DEVS-based M\&S. These needs motivated the proposal of several extensions such as Cellular Automata semantics (CellDEVS \citet{Wainer2002}), parallel semantics (P-DEVS \citet{Chow1994ParallelFormalism}), variable structure (DS-DEVS\citet{Barros1997}, Dyn-DEVS \citet{Uhrmacher2001}), multilevel dynamics (ML-DEVS \citet{Steiniger2016}) just to name a few.  

In our opinion, there was still a need for a DEVS-compliant formalism that specifically addresses  requirements for emergent behaviour modelling. Therefore, in  \citet{Foguelman_2021} we introduced EB-DEVS, a backward compatible extension of Classic DEVS that allows for explicit and direct expression of  macro-level states, upward causation/downward information mechanisms, and live identification of emergent properties.

\subsection{EB-DEVS}

EB-DEVS presents several differences with Classic DEVS that enable a process of top-down and bottom-up information sharing to simulate weak and strong emergence. In EB-DEVS the state transition functions $\dint$ and $\dext$ can access a parent's model state by using the value-coupling $v_{down}$ function. The state transition functions generate new special purpose values to communicate the results to the parent model. We review here the main features of EB-DEVS (see \citet{Foguelman_2021} for a detailed account of all properties).

Formally, an EB-DEVS \emph{atomic} model $M_{\textit{EB-DEVS}}$ is defined by the following structure:
\begin{align}
  M _{\textit{EB-DEVS}}= &\langle \overbrace{X,
    Y,
    S,
    ta,
    \dint,
    \dext,
    \lambda,}^{\text{Classic DEVS}}~\overbrace{Y_{up},
    S_{macro}}^{\text{EB-DEVS extension}}  \rangle
\end{align}

$Y_{up}$ is the set of upward output events (directed to the parent model) and $S_{macro}$ is the set of the parent's states (accessed via downward value coupling). Both sets enable the creation of micro-macro feedback loops. Values $y_{up} \in Y_{up}$  are responsible for triggering upward causation events, causing the invocation of the global state transition function $\delta_G$ at the macro level. Correspondingly, the values $s_{macro} \in S_{macro}$ are copies (or transformations) of macro level states  made available at micro level models.

In Classic DEVS, coupled models are static containers that enable modular and/or hierarchical compositions of other DEVS (coupled and/or atomic) models. However, they have no behaviour nor state of their own. In EB-DEVS both state and dynamics are introduced. Formally, an EB-DEVS \emph{coupled} model $CN_{\textit{EB-DEVS}}$ is defined by the following structure:
\begin{align}
  {CN_{\textit{EB-DEVS}}} & = \langle \overbrace{ X_{self}, 
    Y_{self},D, 
    \{M_i\},
    \{I_i\},
    \{Z_{i, j}\},
  {Select} ,}^{\text{Classic DEVS}}  \overbrace{X^b_{micro}, 
                   Y_{G_{up}},
                   S_{G_{macro}},
                   S_G, 
                   v_{down},
                 \delta_G  }^{\text{EB-DEVS extension}} \rangle
               \end{align}

Upward causation events towards a parent model are defined by the set $X_b^{micro}$. Conversely, information from a parent model is defined by the set $Y_{G_{up}}$. 
A value from the set $S_{G_{macro}}$ is communicated from the parent model. The $CN_{\text{EB-DEVS}}$  state is defined by the set $S_G$, any other value would not be a valid state for this model.
The $v_{down}: S_G \rightarrow S_{macro}$ function transforms and conveys information that lower-level models can consume from the upper-level state.
The global transition function $\delta_G: S_G \times \Re_{0}^{+} \times X^b_{micro} \times S_{Gmacro} \rightarrow S_G \times Y_{Gup} $ computes a new macro state $S_G$ based on its own current state, the elapsed time of the current state, the messages $X^b_{micro}$ arrived from the micro components, and its parent's macro state $S_{Gmacro}$. It also computes the upward causation event (a value in the $Y_{G_{up}}$ set) towards its parent. The cascade of upward causation events can eventually climb up in the hierarchy, possibly (but not necessarily) reaching the topmost (root) coupled model.

\subsection{Synthesis of the relationship between DEVS and EB-DEVS through a minimalist example}

\begin{figure}[htb!]
\centering
\includegraphics[width=1\textwidth]{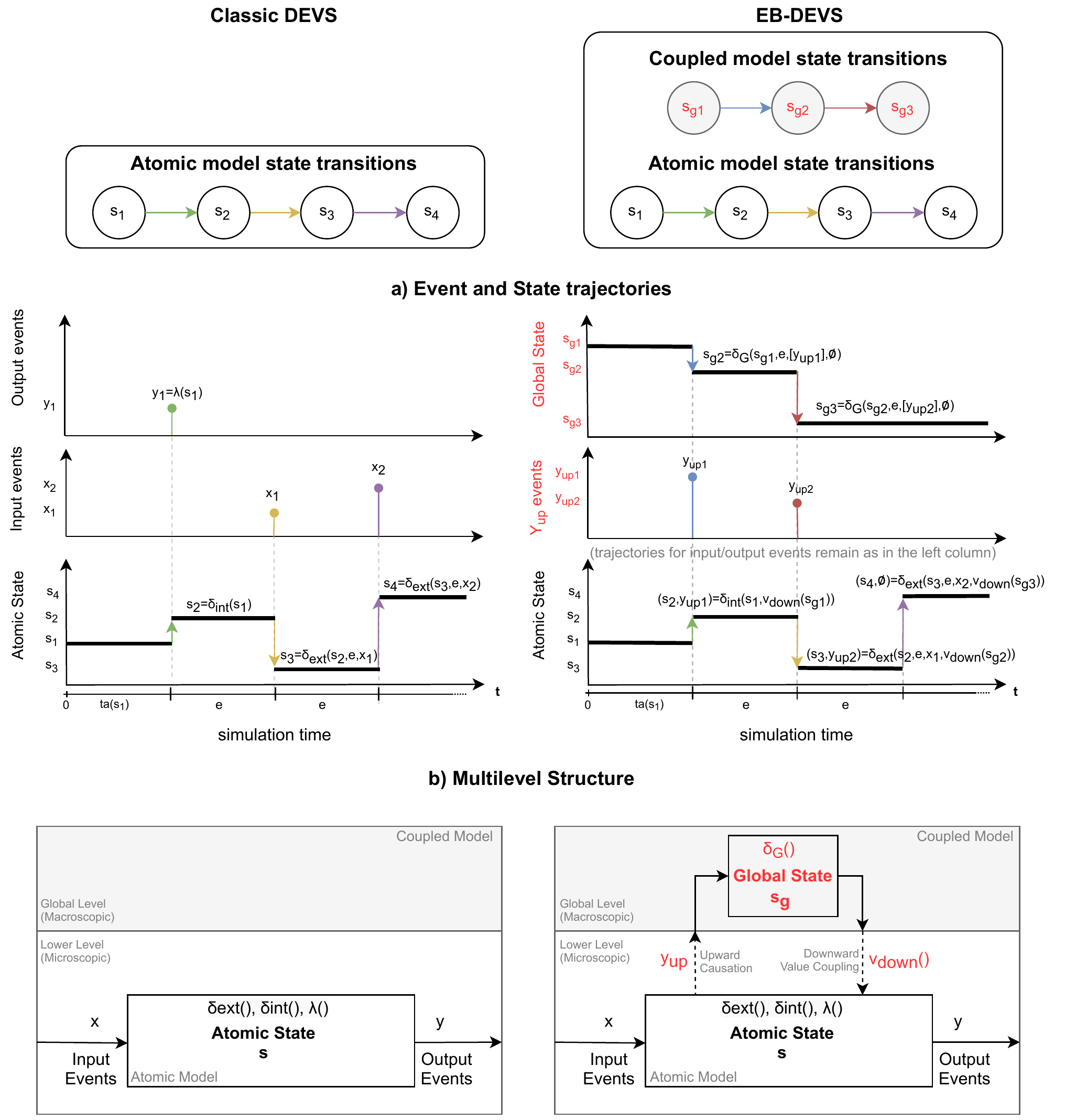}
\caption{A minimalist example of a Classic DEVS model (left column) and its extension into a an EB-DEVS model (right column) showing the new structure and functions involved and a possible global state trajectory and associated state transition diagrams. The example is meant to capture the simplest hierarchical/modular construct conceived in DEVS, i.e. a single coupled model containing a single  atomic model.}
\label{fig:example}
\end{figure}

In \autoref{fig:example} we schematise the core differences and similarities between DEVS and EB-DEVS by modelling a minimalist example system. The columns depict state transitions, main trajectories (panel a) and multilevel relationships (panel b) for the Classic DEVS example (left) and for its EB-DEVS extension (right).

To facilitate the comparison, the same state, input and output trajectories have been defined for the atomic model. As it can be seen in \autoref{fig:example} (panel b, right column) a new global state, global transition function, and multilevel interactions are introduced. Accordingly, the atomic model state transition functions (panel a) extend their signatures to incorporate the micro-macro interactions.

We remark the illustrative nature of this example, as any useful model expected to capture emergent behaviour should have not one, but a (typically large) set of models at the microscopic level for emergence phenomena to occur. For a fully fledged definition and explanation of EB-DEVS and its comparison with Classic DEVS see \cite{Foguelman_2021} including the proofs showing that both formalisms can coexist within a same system, and that both types of models can be incrementally modified to converge on each other.

\subsubsection{A minimalist Classic DEVS model}
\label{example1} 

Consider a minimal system with only one atomic model $m_1$ at state $s_1$. After $ta(s_1)$ units of time $m_1$ adopts the new state value $s_2$ by undergoing an \emph{internal transition} with $s_2=\dint(s_1)$, where $\dint$ ($\dint:S\rightarrow S$) is the \emph{internal transition function}. Just before transitioning from $s_1$ to $s_2$ an \textit{output event} is produced with value $y_1=\lambda (s_1)$, where $\lambda$ $(\lambda:S \rightarrow Y)$ is the \emph{output function}.  %Functions $ta$, $\dint$, and $\lambda$ define the behaviour of a DEVS model.

When an input event $x_1$ arrives through the models' input ports, the external state transition function $\dext$ ($\dext:S\times \Re^+_0 \times X \rightarrow S$) is invoked.
The new state value $s_3$ depends only on the previous state value, the elapsed time since the last transition and the input event value. 

If the model $m_1$ receives an input event at time $t_2+e$ with value $x_1$, the new state is calculated as $s_3=\dext(s_x,e,x_1)$ (note that $ta(s_3) \ge e$). Eventually, the atomic model receives another external message $x_2$ triggering an external transition $s_4=\dext(s_x,e,x_2)$.
No output event is produced during an external transition.

% \begin{figure}[h!]
%     \centering % <-- added
% \begin{subfigure}{0.5\textwidth}
%   \includegraphics[width=\linewidth]{imgs/devs-example.png}
%   \caption{DEVS}
%   \label{fig:example-devs}
% \end{subfigure}\hfil % <-- added
% \begin{subfigure}{0.5\textwidth}
%   \includegraphics[width=\linewidth]{imgs/ebdevs-example.png}
%   \caption{EB-DEVS}  
%   \label{fig:example-ebdevs}
% \end{subfigure}

% \caption{A DEVS example (left), and its extension using EB-DEVS (right) showing the state trajectories corresponding state machines}
% \label{fig:example}
% \end{figure}

\subsubsection{A minimalist EB-DEVS extension}
\label{example2} 

Consider an extension of the previous Classic DEVS model now using the EB-DEVS structure. The atomic model $m_1$ now belongs to the $C_1$ EB-DEVS coupled model that includes the downward information and upward causation mechanisms involving the $\delta_G$ and $v_{down}$ functions. We show a simple example with only two levels and one atomic model to introduce the reader with  the execution process, more elaborated examples will be displayed in \autoref{sec:models}. 

%   \begin{wrapfigure}{r}{0.5\textwidth} 
% %   \vspace{-10pt}
% % \begin{figure}
% %   \begin{center}
%     \includegraphics[width=0.48\textwidth]{imgs/ebdevs-example.png}
%     \caption{An EB-DEVS example showing the state trajectories and state machine.}
%     \label{fig:exampledevs}
% %   \end{center}
% %   \end{figure}
%   \vspace{-20pt}
% \end{wrapfigure}

To begin with, the internal transition of model $m_1$ will execute at time $ta(s_1)$ as before. To do so, it may also take into account the macro-level information $v_{down}(s_{G1})$ using the downward information channel. The internal transition $(s_2, y_{up1}) = \dint(s_1, v_{down}(s_{g1}))$ changes the state of model $m_1$ to $s_2$ and triggers an upward causation event by sending a $y_{up1}$ value. This causes an invocation of the $\delta_G$ function, changing the coupled model's macro-level state to $s_{g2} = \delta_G(s_{g1}, e, [y_{up1}], \emptyset)$. The $\delta_G$ transition parent's macro-state value is $\emptyset$ as there are only two levels in this model. Furthermore, in the example there is only one atomic model thus only one upward causation event will be gathered (characterised with a set of $y_{up1}$ values), but the mailbox collects (potentially) multiple upward causation messages. 

When $m_1$ receives the input event $x_1$, it takes into account the macro-level state provided by $v_{down}(s_{g2})$ and transitions into the new state $(s_3, y_{up2}) = \delta_{ext}(s_2, e, x_1, v_{down}(s_{g2}))$. This state change triggers a new $\delta_G$ transition in $C_1$ generating a new macro-level state $s_{g2} = \delta_G(s_{g1}, e, [y_{up1}], \emptyset)$. 

Finally, the atomic model receives a new input message $x_2$, this time the model will not generate any $y_{up}$ message and therefore the coupled model will remain in its previous global state $s_{g3}$.

\subsection{Relevant features of Complex Systems from the perspective of Emergent Behaviour modelling}

In the modelling of complex systems we identified the need for a systematic and reproducible approach to the identification and incorporation of emergent properties. This need is based on the fact that emergent properties are a common denominator in complex systems involving two or more levels of hierarchy. The introduction of new tools that allow the integration of multiple levels of analysis requires the characterisation of common templates of structural features that relate multiple levels within models.

Complex systems such as the artificial societies reproduced in this manuscript are susceptible to identifying and/or incorporating emergent properties. Therefore, we will focus on typical structural features that we consider key in the modelling of such complex systems with emergent properties. 

%While modelling complex systems, we often find \textit{structural features} that are pervasive in a wide variety of models. 

The subsection converges to \autoref{tbl:models_features} that synthesises where these features can be found in the example models presented later in \autoref{sec:models}.

We will focus on the following structural features: the type of communication structure between the agents, whether the communication channel is explicit or implicit, whether the communication structure is static or dynamic, the types of micro-macro communication channels (top-down information and bottom-up causality), the type of emergence present in the model (weak or strong) and how emergent properties are identified (during the simulation run or a posteriori).

In our opinion, said patterns are often overlooked when designing simulation models, while we consider that a proper identification of such structural features is essential for their clear and correct implementation (helped by clear semantics offered by the modelling formalism of choice).

Therefore, we classify these structural features in five categories, involving relations between agents and modelling levels.

\begin{description}
  \item[Explicit/Implicit communication structure:] When defining the communication mechanism among agents two main types of channels are often used. \textit{Explicit channels} are those where a message is sent from the source agent directly to the destination agent. On the other hand, \textit{implicit channels} can be used for indirect coordination between agents using a stigmergic process \citep{Grasse1959LaConstructeurs}, a coordination mechanism among entities that use their shared environment as a medium for indirect communication. A microscopic entity (agent) can explicitly retrieve updated information from a macroscopic entity (environment) upon necessity.
  
  Both communication strategies are important. On the one hand, explicit messaging is a causal mechanism: when a message is sent (cause) a state change can be observed (effect). On the other hand, implicit communication is useful when the conditions for the information sharing process are more complex. For instance, when agents interact upon proximity, or triggered by an environmental condition like the scent of pheromones that ants leave in their trails, we rely on indirect information and coordination mechanisms (a comprehensive analysis of the impact, role and evolution of direct and indirect communication mechanisms in simulation models can be found in \citet{Tummolini2009StigmergicApproach}).
  
  Explicit channels are built into the DEVS formalism by means of links and input-output ports. EB-DEVS extends the communication structure by adding a downward information channel in the mechanism, thus allowing agents at a lower layer to consume information that is only present at a higher level model and, as a consequence, enabling indirect communication among agents. 
  \item [Static/Dynamic structure:] Dynamic structure models come in a variety of types. The \textit{dynamic} aspect may refer to the communication machinery (as mentioned above), to the ability of adding/removing agents from the system, or to change a model's position in a hierarchy. 
  According to \citet[p.45]{wainer2009}, there are three structural change types that can occur on the system dynamics:
    \begin{enumerate}
        \item System level: changing the relationship between higher-level entities, for instance between coupled models.
        \item Component level: changing the explicit communication network between lower-level atomic models. This happens at the coupled model level.
        \item Subcomponent level: structural changes happen at the atomic level. For instance to stop accepting input events or emitting output events.
    \end{enumerate}
    
    In this work, two example models (Preferential attachment (\autoref{sec:pa}) and SIR (\autoref{sec:sir})) use an implementation of Dyn-DEVS \citep{Uhrmacher2001} that allows for system-level and component-level changes.
    
  \item[Downward information/Upward causation:]
    As for multilevel interactions in a system  \citet{campbell1974downward} and \citet{EMMECHE1997} have discussed how higher levels affect or restrict the course of action of bottom level processes. Downward \textit{causation}, a term coined by Campbell, reflects how a system at the top-level causally drives changes in  systems at bottom levels. Analogously, the downward \textit{information} mechanism only presents higher-level information to lower-level processes, while this information can be either used or dismissed. Upward causation can be seen as a reciprocal, bottom-up relationship. %Apparently, these two notions cannot be separated. Mario Bunge's treaty on causality \cite{bunge2017causality} (p.362) exposes both relationships and criticizes the use of the term `causation' in this context, as he states that ``\textit{what we do have here is not causal relations but functional, relations among properties and laws at different levels}''.
    
    EB-DEVS implements upward causation by using $y_{up}$ events. When a lower-level model stores a value in the $y_{up}$ variable, an upward causation event will be triggered at its parent model. As for the downward information channel, lower-level models have access to the macro-level state information through the $v_{down}$ function.
    \label{item:updown}

  \item [Weak/Strong Emergence:] In this work we adopt the definitions in \citet[p.4]{Szabo2017}: ``\textit{weak emergence as being the macro-level behaviour that is a result of micro-level component interactions, and strong emergence as the macro-level feedback or causation on the micro-level}''.
  To represent weak emergence with EB-DEVS we resort to the upward causation mechanism, and to model strong emergence we include the downward channel bringing macro-level information to the micro-level models. This closes the required micro-macro feedback loop allowing for adaptive strong emergence (see \citet{Mittal2013} for a discussion of relations between the DEVS formalism, Emergence, Stigmergy and Complex Adaptive Systems modelling).

  \item[Emergence identification:] Emergent properties can be modelled and/or identified. Identification can happen either during the simulation (live identification) or after the simulation has finished (ex-post identification).
  
  In the case of live identification, emergent properties are detected at higher-level entities (such as coupled EB-DEVS models). 
  
  EB-DEVS features a macro-level state $s_G$  for each coupled model that is key to perform live identification  by accessing the variable (or variables) that hold information on components, states and aggregated interactions (this is consistent with the concept of Emergence Behaviour Observer snapshots in \citet{mittal2015harnessing}).
 
  Ex-post identification of emergent properties are detected while analysing the results of a simulated model and usually does not intervene in the modelling process, while live identification allows for modelling behaviours that depend on the identified properties.

\end{description}

In \autoref{tbl:models_features} we synthesise the structural features for each of the models that we develop throughout the rest of the paper.
Besides the \textit{structural features} discussed above there are two salient \textit{Properties} that we consider worth specifying (also included in \autoref{tbl:models_features}).

\begin{table}[htb!]
\small
\begin{tabular}{|p{0.20\linewidth}|L|L|L|c|c|}
\hline
\textbf{Structural Feature}   & \textbf{Dissemination of Culture} & \textbf{Segregation} & \textbf{Preferential Attachment}  & \textbf{Sugarscape} & \textbf{SIR} \\ \hline
Explicit/Implicit communication structure & E          & I            & E& I/E  & E   \\ \hline
Static/Dynamic structure      & S          & S            & D& S & D \\ \hline
Downward information/ Upward causation   & D+U        & D+U & U           & D+U    & D+U \\ \hline
Weak/Strong Emergence         & S          & W   & W         & S    & S   \\ \hline
Emergence identification      & Live       & Ex Post        & Ex Post            & Live & Live    \\ \hline
\textbf{Properties}&&     &           &  &         \\ \hline
Macroscopic Level Algorithm   & CL         & None& PA   & Gini index     & PA               \\ \hline
Topology Type  & Lattice / Arbitrary   & Lattice        & Scale Free    & Lattice & Scale Free \\ \hline
\end{tabular}
\\

\caption{Models implementing emergent characteristics.}
\label{tbl:models_features}
\end{table}

The \textbf{Macroscopic Level Algorithm} refers to the technique used to model macro-level aggregate states. During the implementation of our example models we adopted form simple aggregations and clustering techniques  to more complex algorithms like size-biased sampling, Preferential Attachment or Gini index calculation.

The \textbf{Topology Type} refers to the structures defined to communicate sub components within a model. In this work we used three types of communication graphs: lattice, random, and scale-free graphs. The implemented models use an input graph that defines the initial connectivity between input and output ports.

In the next section we present five simulation models involving  the aforementioned features and properties, and discuss the modelling of emergent behaviour in the context of EB-DEVS.

% \subsection{Classic DEVS and EB-DEVS}
% \subsubsection{Illustrative simple example}

\section{Modelling Emergent Behaviour in Social Dynamics with EB-DEVS}
\label{sec:models}

In this section we describe the models implemented. We provide for each of them a general description followed by an analysis regarding their features related to complex systems and emergent behaviour modelling. For an exemplification of the EB-DEVS implementation of different dynamics we outline the pseudocode for each case (adopting a colour convention to highlight the upward causation and downward information mechanisms, and a syntax similar to that of the Python language). 
Finally, the experimental results for each case study are presented, including a brief discussion regarding the modelling experience  using an emergent behaviour-explicit formalism.

\subsection{Dissemination of Culture with strong emergent properties}
\label{sec:axelrod}

Axelrod's model of Dissemination of Culture has attracted considerable attention because of its simple rules and its implications for social dynamics. Its main interest lies in the mechanisms it implements to model cultural assimilation. The model implements two important features: social influence and homophily. A social influence model is expected to converge to a monocultural equilibrium. However, Axelrod's model presents scenarios in which the system moves away from monocultural towards multicultural equilibrium (\cite{Centola2007HomophilyGroups}). These scenarios are often determined by the number of features and traits that define each culture.

\subsubsection{Model dynamics}

Each agent has a state with two variables: a features array and a dictionary with the neighbours' features arrays. The feature array has $F$ values of cultural traits and each cultural trait is defined in the range of integers $[1,Q]$.
The state of agents change as they interact with neighbours (assuming an arbitrary topology). An agent selects a neighbour to mix its culture with. and decides to interact based on its similarity. In this context, similarity is defined as the average number of matching traits (i.e. the number of matching traits divided by $F$). If their similarity is greater than a randomly chosen value, cultural mixing occurs. The process consists of choosing, for each agent, one of the different traits in the set of cultural traits and agreeing on a common value with the neighbour.
 
This simple model exhibits interesting behaviours and has seen several extensions pursuing different degrees of sophistication \citep{Castellano2000NonequilibriumInfluence,Balenzuela2015, Klemm2003GlobalSystems}. From a modelling perspective, the model presents adaptive behaviour (as opposed to rational behaviour) without central authority.

We show in \autoref{lst:axelrod} an implementation of the basic dynamics for the culture diffusion model. We provide the pseudocode for a Classical DEVS atomic model representing an agent within a group of interacting agents contained in a same Classical DEVS coupled model (representing the society as a whole).

\begin{lstlisting}[style=base,
    label={lst:axelrod},
    caption={Pseudocode for the Axelrod's Dissemination of Culture model in Classic DEVS.}]
AgentState:
  neighbours_cultures = {neighbour_id: culture}
  self_culture = [randint(1, Q) for x in range(F)]

state = AgentState()
AgentAtomicInternalTransition():
  # Get a random neighbour culture from the ones it knows.
  random_neighbour = random_choice(state.neighbours_cultures.keys())
  # Find the matching positions in the culture array
  matching_cultures = get_matching_cultures(random_neighbour)
  share_culture = False
  # Toss a coin weighted by the percentage of matching cultures
  if random() < matching_cultures / F:
    # Randomly choose a trait that is different for both agents and copy it.
    non_matching_cultures = get_non_matching_cultures(random_neighbour)
    random_culture, random_culture_index = random_choice(non_matching_cultures)
    state.self_culture[random_culture_index] = random_culture
    share_culture = True
    
AgentAtomicExternalTransition(neighbour_id, culture):
  # Update local neighbours information
  state.neighbours_cultures.update({neighbour_id: culture})
  
AgentAtomicOutputFunction():
  # Share agents culture with neighbours.
  share_with_neighbours(state.self_culture)
  
AgentAtomicTimeAdvance():
  # Time advance function, if sharing then share the culture immediately. 
  if share_culture:
    return 0
  else:
    return 1
\end{lstlisting}

\subsubsection{Emergence and adaptive behaviour: A fads and fashion scenario}

The capabilities of EB-DEVS become relevant when considering extensions to the standard Axelrod's model. Let's examine the case of a \textit{Fads and fashion} scenario. We provide this extension to show how the micro and macro levels can be integrated. 

The concept of \textit{fashion} can be defined as the cultural trait desired by many but defined by few. We can consider that the fashion for a cultural trait is the most popular value among the agents of the system. Said trait can only be observed from a macroscopic level. When this information is made available to the system it can then influence the evolution of agents' culture over time.

We extend the original model to include this idea in the following way. At the macroscopic level (coupled model) we include a stochastic variable that yields fashion trait values from the agents' culture arrays. First, it generates a random value in the range $[1,F]$ and then it selects the $q \in Q$ that is most frequent in the system. This type of event (from now on a \textit{fashion cultural mix} event) is produced at a  rate defined by a parameter. During the internal transition an atomic model (i.e, an agent) will toss a weighted coin ($P=FASHION\_RATE$) to undergo either a \textit{fashion cultural mix} or a \textit{regular cultural mix}. At the limit values of FASHION\_RATE, the model behaves either as the classic model ($P=0$) or as a fashion doppelganger model ($P=1$).

In \autoref{lst:axelrod_fashion} we show the pseudocode of the \textit{Fads and fashion} extension.

\begin{lstlisting}[style=base,
    label={lst:axelrod_fashion},
    caption={Pseudocode of the Dissemination of Culture model with fads and fashion using EB-DEVS. Micro-macro interactions are highlighted with red for $v_{down}$ and blue for $y_{up}$.}]
AgentState:
  neighbours_cultures = {neighbour_id: culture}
  self_culture = [randint(1, Q) for x in range(F)]

state = AgentState()
AgentAtomicInternalTransition():
  # Get a random neighbour culture.
  random_neighbour = random_choice(state.neighbours_cultures.keys())
  # Find the matching positions in the culture array.
  matching_cultures = get_matching_cultures(random_neighbour)
  share_culture = False
  # Toss two weighted coins: first determine if it should do Fads and Fashion, then toss a coin weighted by the percentage of matching cultures. Events are mutually exclusive.
  do_fashion = random() < FASHION_RATE
  if not do_fashion and 
          random() < (matching_cultures / F):
    # Randomly choose a trait that is different for both agents and copy it.
    non_matching_cultures = get_non_matching_cultures(random_neighbour)
    random_culture, random_culture_index = random_choice(non_matching_cultures)
    state.self_culture[random_culture_index] = random_culture
    share_culture = True
  else:
    # Ask the parent model for a fashion cultural trait.
    @feature, fashion = @$\color{red}v_{down}(FASHION\_FEATURE)$
    state.self_culture[feature] = fashion
  # Upward causation message, triggering $\color{red}\delta_G$ transition.
  /*$\color{blue}y_{up}$ = (state.agent_id, state.culture)/*

AgentAtomicExternalTransition():
  # Update local neighbours information.
  state.neighbours_cultures.update({neighbour_id: culture})

AgentAtomicOutputFunction():
  # Share agent's culture with neighbours.
  share_self_culture_with_neighbours()

AgentAtomicTimeAdvance():
  # Time advance function, if sharing the state then share the culture immediately.
  if share_culture:
    return 0
  else:
    return 1

CoupledState:
  models_cultures = {}

$\color{red}s_{G}$ = CoupledState()
CoupledGlobalTransition($e_g$, $\color{blue}x^b_{micro}$, $\color{red}s_{Gmacro}$):
  # Update the global status with the cultures from each model
  $\color{red}s_{G}$.models_cultures.update($\color{blue}x^b_{micro}$)
@  
$\color{red}v_{down}$(PROPERTY)@:
  if PROPERTY == FASHION_FEATURE:
    cultures_matrix = $\color{red}s_{G}$.models_cultures.values()
    # Pick a random cultural trait.
    randfeature = randint(F)
    feature = cultures_matrix[:, randfeature]
    # Get the most frequent trait value.
    fashion_value = bincount(feature).argmax()
    return randfeature, fashion_value@
\end{lstlisting}

\subsubsection{Experimental results and discussion} 

\begin{figure}[htb]
    \centering % <-- added
\begin{subfigure}{0.5\textwidth}
  \includegraphics[width=\linewidth]{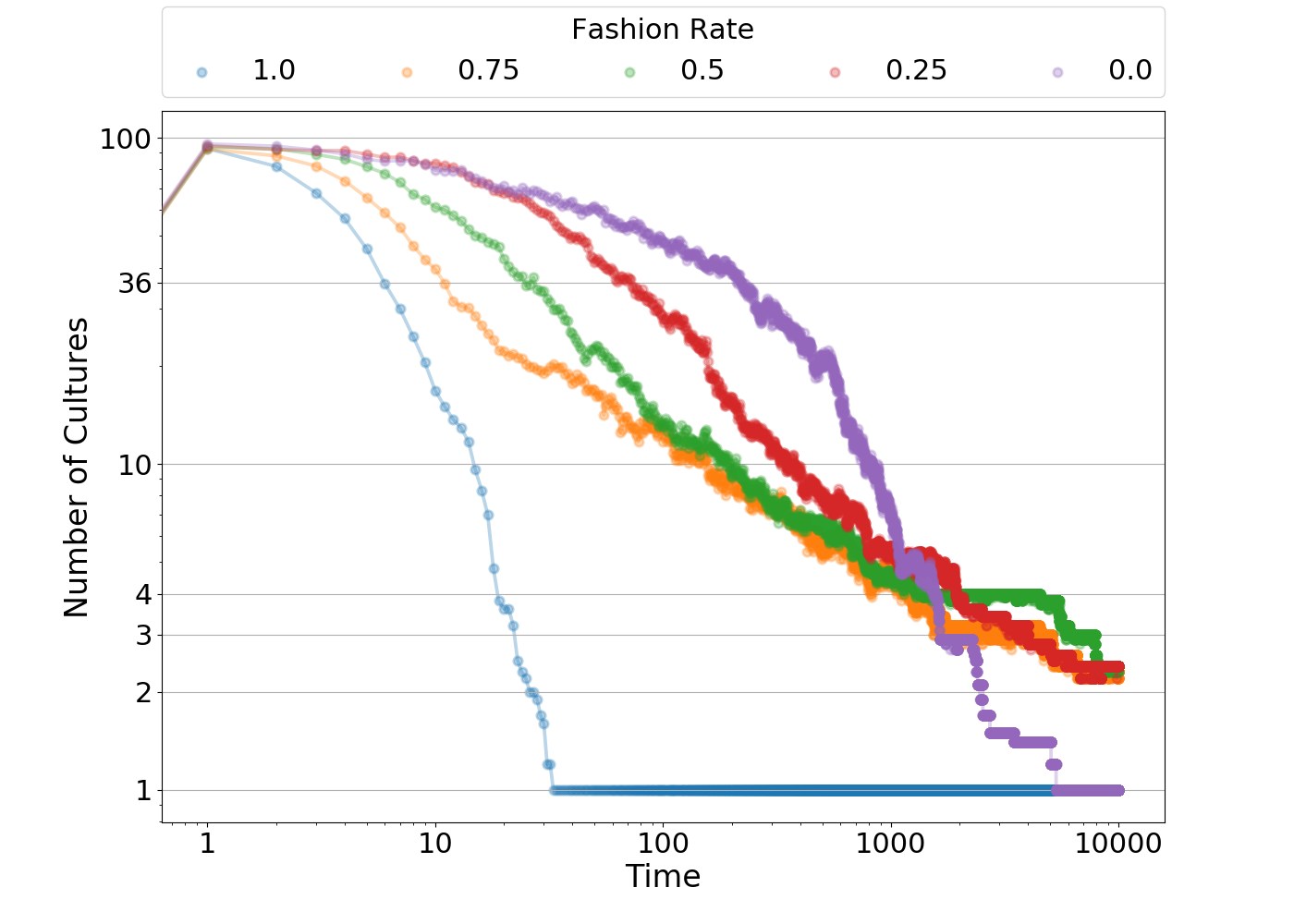}

  \caption{Dissemination of Culture for F=5, Q=5.}
  \label{fig:axelrod_1}
\end{subfigure}\hfill % <-- added
\begin{subfigure}{0.5\textwidth}
  \includegraphics[width=\linewidth]{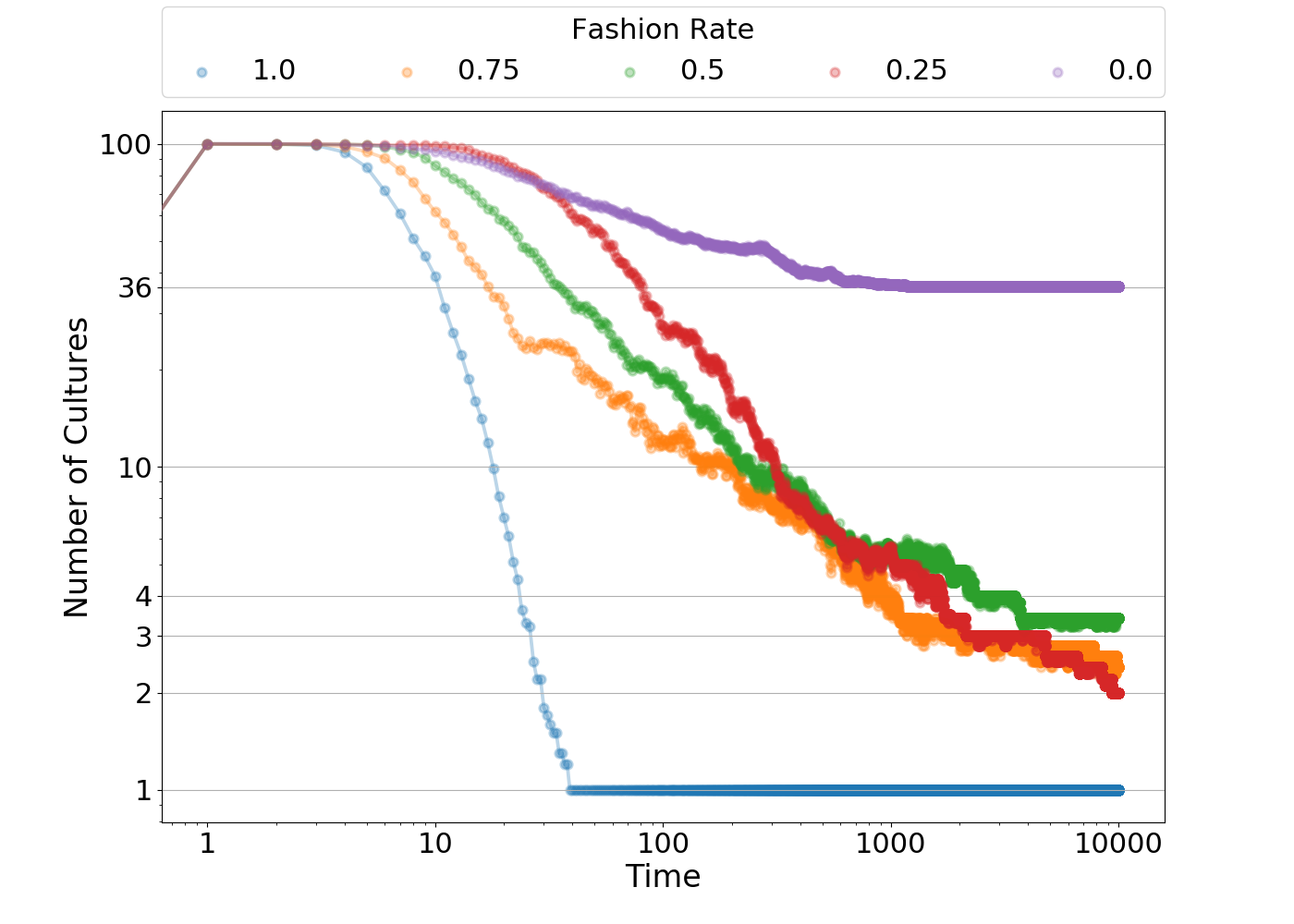}
  \caption{Dissemination of Culture for F=5, Q=20.}
  \label{fig:axelrod_2}
\end{subfigure}

\caption{Dissemination of Culture model for 5 cultural features (F), different fashion rates and different numbers of possible cultural traits (Q). The original model (FASHION\_RATE = 0, no fashion ever considered) shows a convergence to a monocultural system (left, Q=5) vs. a highly scattered multicultural system (right, Q=20). On the other end (FASHION\_RATE = 1, fashion always considered) there is a consistent convergence to monoculture. At intermediate levels of FASHION\_RATE the system converges to a slightly scattered multiculture.}
\label{fig:axelrod}
\end{figure}

The experiments in \autoref{fig:axelrod} were configured as follows: $F=5$, $Q=\{5, 20\}$ and $N=100$. We define $N$ agents as atomic models within a same coupled model. The connectivity network is a 10x10 regular lattice. Each simulation realisation is run for 10000 time units.

The parameters $F$ and $Q$ were selected to show two different scenarios of convergence for the classic model, i.e. with FASHION\_RATE=0. The scenario with $F=5$ and $Q=5$ shows the model converging towards a monocultural final state, while the scenario with parameters $F=5$ and $Q=20$ converges towards a multicultural state. This is consistent with the original model where small values of $F$ present monocultural convergence and larger values of $Q$ present multicultural convergence. We want to study how the FASHION\_RATE affects the system evolution and convergence.

We swept the FASHION\_RATE parameter from 0 to 1 with intervals of 0.25, with 10 realisations per experiment. The plots in \autoref{fig:axelrod} show the evolution of the mean number of cultures through time.

In \autoref{fig:axelrod_1} we can observe that the experiments converge towards a monocultural system for FASHION\_RATE = 0. This, again, is consistent with the original model. Nevertheless, we observe that with FASHION\_RATE=1 the model also converges towards a monocultural system although at a faster rate. Furthermore, the scenario is clearly affected by intermediate values of FASHION\_RATE = \{0.75, 0.5, 0.25\} where the system now converges towards multicultural states (with a  mean number of cultures in the range of 2 to 3)

Looking at \autoref{fig:axelrod_2}, we see that the original model (FASHION\_RATE = 0) converges to a multicultural system with an average of 36 cultures. Interestingly, for FASHION\_RATE = 1 the model converges to a monocultural system and for values of FASHION\_RATE = {0.75, 0.5, 0.25} the system converges again to a multicultural system with few cultures.

% It is noteworthy how the fashion rate affects the cultural convergence, where it would converge to a high number of cultures fashion affects and reduces this number towards a monocultural system \autoref{fig:axelrod_2}.

The Dissemination of Culture model's implementation in Classic DEVS is simple. From the modelling perspective, the utilisation of explicit communication channels enables the culture propagation with clear semantics. While the DEVS M\&S framework resolves the communication protocol between agents, emergence identification can only be done ex-post, i.e. by analysing the logged time series generated by the experiments. 

In \autoref{fig:axelrod} we observe a stable macroscopic pattern generated by the agents' interactions (this observation has been discussed in the computational social sciences field, see \citet[p.33]{Epstein1999}). This emergent property cannot be detected at simulation time using Classic DEVS. 

%Thus, we find that DEVS is insufficient for the expression of live emergent behaviour that EB-DEVS is capable of.

% The implementation of the Dissemination of Culture model using DEVS is simple and the benefits of using DEVS as the M\&S formalism are many. The simulation execution and scheduling, the integrated messaging protocols, and the structured modelling hierarchies improves the programming effort diminishing development time. 

% This is easily seen in the rather simple pseudo-code shown in \autoref{lst:axelrod}. Yet, model's emergence identification can be only done ex-post using statistical time series methods to detect \textit{`a stable macroscopic or aggregate pattern induced by the local interaction of the agents'} using \citet[p.33]{Epstein1999} words. Furthermore, for live emergent behaviour DEVS lacks from the primitives needed for such task. 

The \textit{fads and fashion} model extends the Axelrod's basic setting by introducing a micro-macro feedback loop to capture  emergent properties. In the model, the system's overall cultural trend is influenced by the interactions at multiple levels. In the \autoref{fig:axelrod} we can see how the fashion rate influences the convergence in two different scenarios. %the system in two ways. In the monocultural instance of the model, accelerates the convergence process. Reciprocally, in the multicultural instance of the model it changes the phase of the model from a multicultural towards monocultural convergence.
We consider this example as a common pattern for the implementation of strong emergence in simulation models. 

Regarding the modelling advantages that EB-DEVS offers for this case study, from an agent's perspective, the upward causation mechanism ($y_{up}$ messages) allows for a seamless integration with macroscopic variables, while the downward information mechanism makes available required macroscopic information demanding very few changes ($v_{down}$ function). EB-DEVS facilitated the inclusion of emergent properties in an expressive and uncomplicated manner into the Dissemination of Culture model.

% Firstly, the utilisation of explicit static communication structures, allows for the information sharing process using EB-DEVS input/output ports. Secondly, the usage of upward causation allows for the calculation of the cultural groups at the upper levels, enabling the downward information mechanism that is used for the implementation of the  \textit{Fads and Fashion} model. 
% Finally, extending the classic Dissemination of Culture with strong emergence behaviour shows how easily new characteristics can be added to any given model.

\subsection{Implicit communication and weak emergence in the Segregation model}

In the 70s Thomas Schelling proposed an agent based model to analyse racial segregation \citep{Schelling1971}. In the model agents are assigned with a colour and an initial position on a grid, where they move according to their \emph{happiness} evaluated as a function of the colours found in the agent's neighbourhood. If an agent is surrounded by neighbours of different colour exceeding a given threshold, then the agent is considered \emph{unhappy} and moves to a random empty position in the grid. Each agent will change its position with a constant time step until it reaches a \textit{happy} state. The simulation converges to a steady state when all the agents are happy, otherwise the simulation continues forever. Schelling used this model to study the formation of areas segregated by the colour of the skin or race of their population \citet[p.145]{Schelling1971}.% (continuous areas with agents of identical colour).

% When the simulation begins, agents start to move to a randomly selected empty cell in the grid. This random movement continues until the agents find a \emph{happy} position. If they do not find such position, they  continue moving without reaching a final stable state.

\subsubsection{Model dynamics}

We model agents as EB-DEVS atomic models with a state defined by two variables: a colour and a position in the grid, which are randomly selected at the beginning of each simulation. Agents are contained by a single EB-DEVS coupled model that manages the grid representing the environment. In particular, the coupled model provides the atomic models with macro-level information such as an available free position where to move or the number of neighbours matching the agent's colour.

Agents do not know about the grid nor the number of agents (the size of the system). Therefore, they resort to the macro-level information to calculate their \emph{happiness} relative to the neighbours colours. In the case of an unhappy result they move to a random empty cell, which is also provided by the coupled model. We utilised the upward causation and downward information capabilities of EB-DEVS for the implementation of indirect communication.% through the environment between the agents and the grid. 
The upward causation event $y_{up}$, triggers the global transition function $\delta_G$ updating the colours in the grid, and the $v_{down}$ function provides a (partially) macroscopic information about the happiness of neighbouring agents.

The model has the following parameters: the size of the grid, the number of agents, and the happiness threshold. In \autoref{lst:schilling} we show the pseudocode of the  model.

\begin{lstlisting}[float,style=base,
    label={lst:schilling},
    caption={Pseudocode of the Schelling's segregation model. Micro-macro interactions are highlighted in red for $v_{down}$ and blue for $y_{up}$ and $x^b_{micro}$.} 
    ]
AgentState:
  # Initial agent's state
  colour = get_random_colour(RED, GREEN)
  position = get_initial_position()
  
state = AgentState()
AgentAtomicInternalTransition():
  # Get information from the parent model regarding its happiness relative to its neighbours.
  @happy = $\color{red}v_{down}$(HAPPINESS, state.position, state.colour)@
  last_pos = state.position
  if not happy:
    # Unhappy agents move to a random empty position defined by the parent model.
    @empty_cell = $\color{red}v_{down}$(RANDOM_EMPTY_CELL)@
    state.position = empty_cell
    /*$\color{blue}y_{up}$ = (last_pos, state.position)/*

AgentAtomicTimeAdvance():
  # Sampled time from the exponential distribution.
  return random_exponential(0.5)

CoupledState:
  # Initial global state of each agent at each position in the grid
  grid = GRID(20,20)

@$\color{red}s_G$@ = CoupledState()
CoupledGlobalTransition($e_g$, $\color{blue}x^b_{micro}$, $\color{red}s_{Gmacro}$):
  # Update the grid position value of each agent communicating its new colour.
  for last_pos, actual_pos in /*$\color{blue}x^b_{micro}$/*:
    colour_swap = @$\color{red}s_G$@.grid[last_pos]
    @$\color{red}s_G$@.grid[last_pos] = EMPTY
    @$\color{red}s_G$@.grid[actual_pos] = colour_swap

@$\color{red}v_{down}$(PROPERTY, parameters):
  # Value Coupling function that communicates a random empty cell and relative happiness value.
  if PROPERTY ==  RANDOM_EMPTY_CELL:
    return random_empty_cell()
  elif PROPERTY ==  HAPPINESS:
    return happiness_from_neighbours(parameters.position, parameters.colour)@
\end{lstlisting}

\subsubsection{Adaptive behaviour and implicit communication}

The implementation of the model presents relevant features from the modelling standpoint. The environment (coupled model) maintains an updated state of the occupancy on the grid, enabling the implementation of implicit communication for proximity-based interactions between agents. This allows to share the happiness and the available positions in the grid with the agents without the need of explicit links between atomic models.

Combining static structures (i.e., the grid) with implicit communication mechanisms enables macroscopic-driven adaptive behaviour at the microscopic-level. Furthermore, this short example provides the basis for weak emergence modelling.

\subsubsection{Experimental results and discussion} 

\begin{figure}[hbt!]
\centering
    \begin{subfigure}{0.45\textwidth}
    \includegraphics[width=\textwidth]{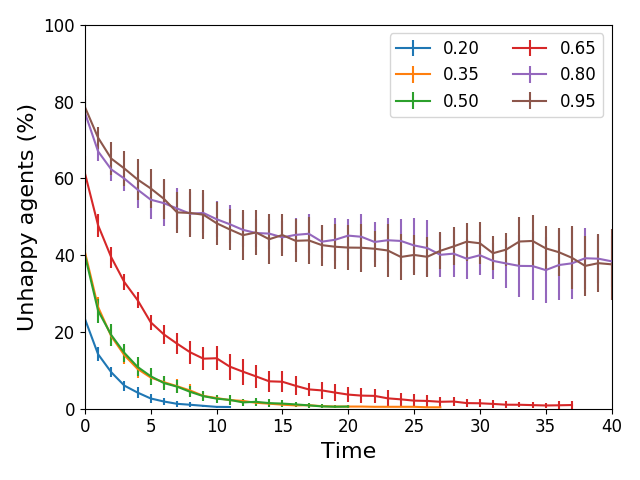}
    \end{subfigure}
        \begin{subfigure}{0.45\textwidth}
    \includegraphics[width=\textwidth]{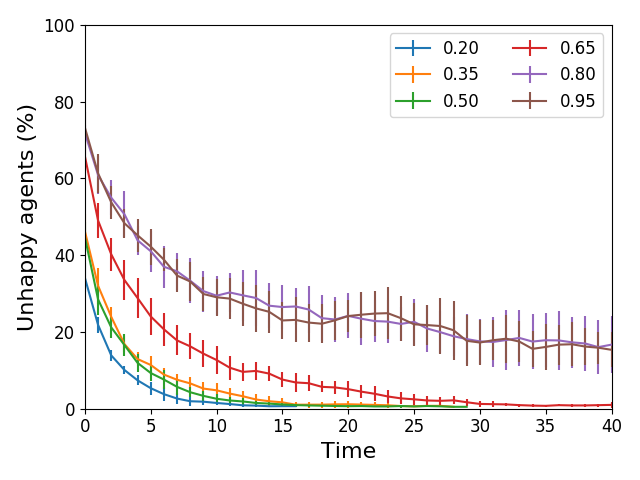}
    \end{subfigure}
    \caption{{Simulation and modelling of the Schelling's segregation model.} Scenarios varying happiness thresholds were simulated for with two different populations: left) 266 agents and right) 200 agents.}
        \label{fig:schelling}
\end{figure}

In \autoref{fig:schelling} we can see the simulation results for the model. The experiments were done for two different population sizes, N=\{266, 200\}. The Happiness Threshold was swept for the values HT = \{0.20, 0.35, 0.50, 0.65, 0.80, 0.95\}. The size of the grid is 20x20 cells and the experiments were configured for a duration of 40 units of time, performing 10 realisations for each combination of parameters.

The system converges when the number of unhappy agents drops to zero. In the cases where the system converges, ghettos appear in the grid as clusters of agents sharing the same colour. System's convergence depends mainly on the size of the system and the happiness threshold. This can be seen analysing the percentage of unhappy agents as a function of time (\autoref{fig:schelling}). 
The cases where the system does not converge to zero unhappy (HT = \{0.8, 0.95\}) were cut off at 40 time units.

%We simulated different scenarios (10 runs for each scenario) varying the happiness threshold and the size of the system.

% In \autoref{fig:schelling} we can see that as the happiness  increases the simulations took longer to finish and there is a certain limit beyond it does not converge. Also, when we changed the size of the system to less dense configurations, we obtained the same behaviour but in not converged cases, the percentage of unhappy agents reaches a different value. These differences are due to the different configuration of parameters, some of them, related to agents behaviour and other to the size of the system.

In this example, we have used implicit communication with static structure. The downward and upward mechanisms, are used for the implementation of indirect communication of the neighbouring happiness values. This implementation reduces the model's complexity in two ways. On the one hand, it reduces the number of  input/output explicit links (with all their associated DEVS message handling overhead) that would otherwise have been required. On the other hand, it allows to update the agents' (implicit) communication structure in an easy way (avoiding to rewire a classic explicit DEVS structure, which requires more sophisticated considerations and mechanisms such as dynamic structure  capabilities (\citet{Uhrmacher2001}).

Finally, we remark that this model exhibits a weak form of emergence, presenting a stable macroscopic pattern allowing only for ex-post emergence identification. It could be possible to extend the model by incorporating strong emergence, which can be considered in a future work.

\subsection{Dynamic explicit networks in the Preferential Attachment model}
\label{sec:pa}

The scale-free property \citep{Albert2002StatisticalNetworks} appears in many complex networks of agents describing social interactions and collaborations, biological relations, the World Wide Web, financial and payment networks, semantics, airlines, just to name a few. The degree of interconnection between agents, i.e.: the degree of each node in the connectivity graph, in these large networks follows a power law distribution, which seems to be a consequence of the expansion of the network following a \textit{Preferential Attachment (PA)} process \citep[p. 76]{Albert2002StatisticalNetworks}. Namely, each newly added node is more likely to connect with existing nodes with high degree. 
Eventually as the system evolves, the nodes start forming hubs that exhibit a power-law distribution in the degree of the nodes. This macroscopic structure originates from microscopic rules of connectivity, and is presented as a form of emergent property of the system.

\subsubsection{Model dynamics}
In order to simulate the Preferential Attachment process in EB-DEVS, we will model the evolving graph as nodes (represented by agents) that attach to the existing network.

We consider an initial graph with two nodes connected by an edge, where each node is implemented as an EB-DEVS atomic model, and edges are explicit links connecting input output ports. At each step of the simulation a new node is added to the network, and is connected with $n$  of the previously existing nodes, which are selected using a random variable sampled from a size-biased distribution, modelled using the coupled model's macro-level state. Therefore, the probability  to select a neighbour of degree $k$ is ${(k. p_k)}/{\sum_{j=1,...,n}(j.p_j)}$ where $p_k$ is the proportion of existing nodes with degree $k$. Following this property, the probability of a node to be chosen is proportional to its degree, making the higher-degree nodes prone to gain more neighbours. Furthermore, older nodes will correlate with hubs. 

A network that evolves following these rules will be composed of nodes whose degree have a scale-free distribution \citep{Albert2002StatisticalNetworks}. The probability $p_k$ that a vertex in the network has degree $k$ follows a power-law distribution. Namely, $p_k$ is proportional to  $k^{-\gamma}$ for some $\gamma$.

\subsubsection{Scale-free network generation and adaptive behaviour} 

The implementation of this model presents several modelling challenges that we consider of interest. As the model requires to dynamically add new atomic components and links, we face a case of dynamic structure in the context of EB-DEVS. Dynamic structure models enable the expression of more complex systems as we will see in \autoref{sec:sir}. Secondly, our preferential attachment implementation resorts to upward causation and weak emergence to drive the evolution of the dynamic structure network. This allows for the emergence of a higher-level structure as it was previously described. This example shows how dynamic networks with explicit links can be built with EB-DEVS in a compact and sound way.

In \autoref{lst:preferential} we show the pseudocode of the Preferential Attachment model.

\begin{lstlisting}[style=base,
  label={lst:preferential},
  caption={Pseudocode of the Preferential Attachment model. Micro-macro interactions are highlighted with red for $v_{down}$ and blue for $y_{up}$ and $x^b_{micro}$.}]
AgentState:
  id = incremental id
  outdegree = 0

state = AgentState()
NodeAtomicInternalTransition():
  # Triggers an upward causation event towards the coupled model carrying its out degree.
  $\color{red}y_{up}$ @= (id, outdegree)@

CoupledState:
  # Model's initial topology
  topology = Parameters.INITIAL_TOPOLOGY
  # Model's state with nodes degrees
  nodes_degree = {}

$\color{red} s_G$ = CoupledState()
CoupledGlobalTransition($e_g$, $\color{blue}x^b_{micro}$, $\color{red}s_{Gmacro}$):
  # Updates the macro level state with the agent's out degree
  for elem in /*$\color{blue}x^b_{micro}$/*:
    nodes_degree[elem.id] = elem.outdegree
  # Defines the dynamic network behaviour:
  # Retrieve the node's degree sequence
  degree_sequence = nodes_degree.values()
  # Create size-biased probability array
  pk = [degree / sum(degree_sequence) for degree in degree_sequence]
  # Sample from the weighted distribution the nodes to connect to
  connect_to_node = choice(nodes_degree.keys(), pk, replace=False, number=Parameters.CONNECT_TO)
  atomic = create_atomic_model()
  # Create, connect and update topology
  for node in connect_to_node:
    connect_models(atomic, node)
  update_topology()
\end{lstlisting}

\subsubsection{Experimental results and discussion} 

\begin{figure}[htb!]
    \centering % <-- added
% \begin{subfigure}{0.5\textwidth}
%   \includegraphics[width=\linewidth]{imgs/preferential_attachment_degree_histogram_power_fit_connect_1}
%   \caption{Connect to one node model.}
%   \label{fig:PA1}
% \end{subfigure}\hfil % <-- added
\begin{subfigure}{0.49\textwidth}
  \includegraphics[width=\linewidth]{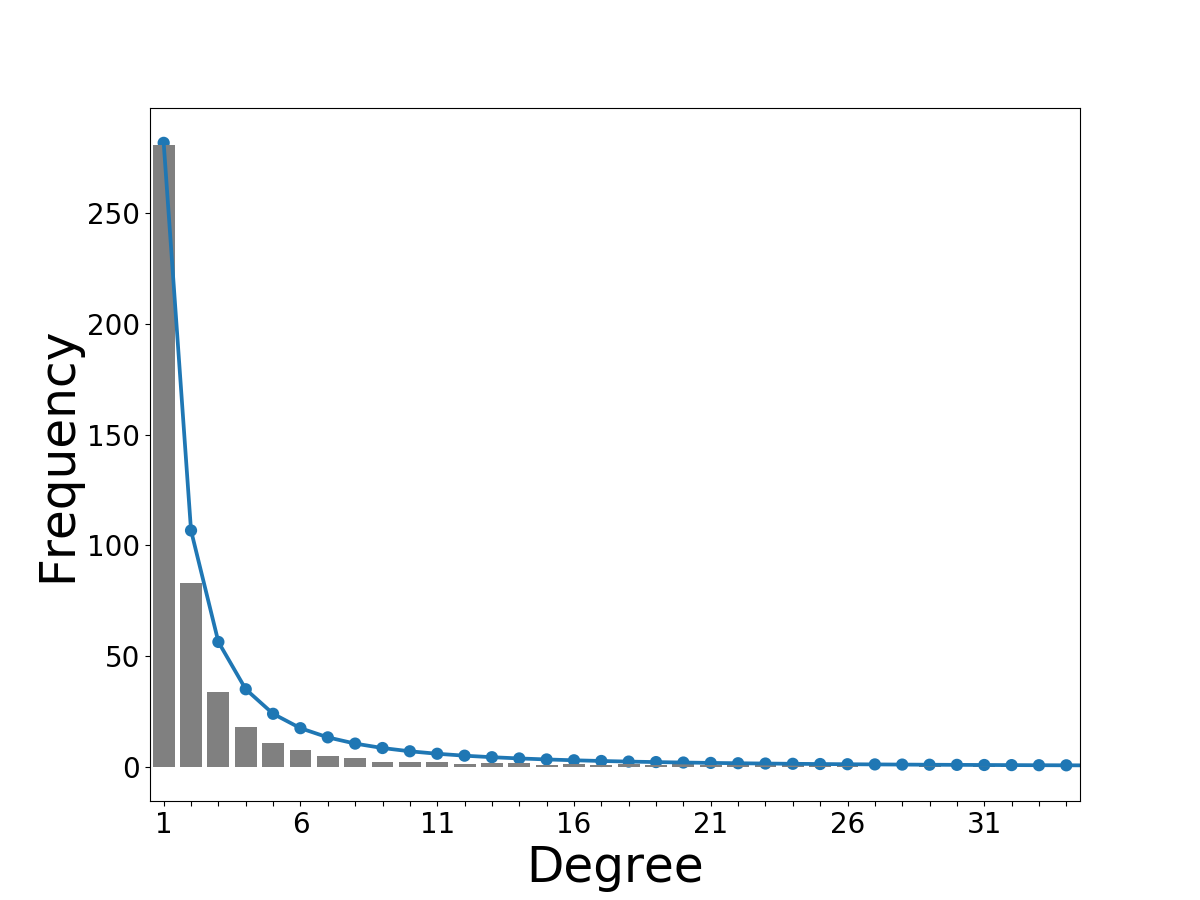}
  \caption{Connect to one node.}
  \label{fig:PA1}
\end{subfigure}
\begin{subfigure}{0.49\textwidth}
  \includegraphics[width=\linewidth]{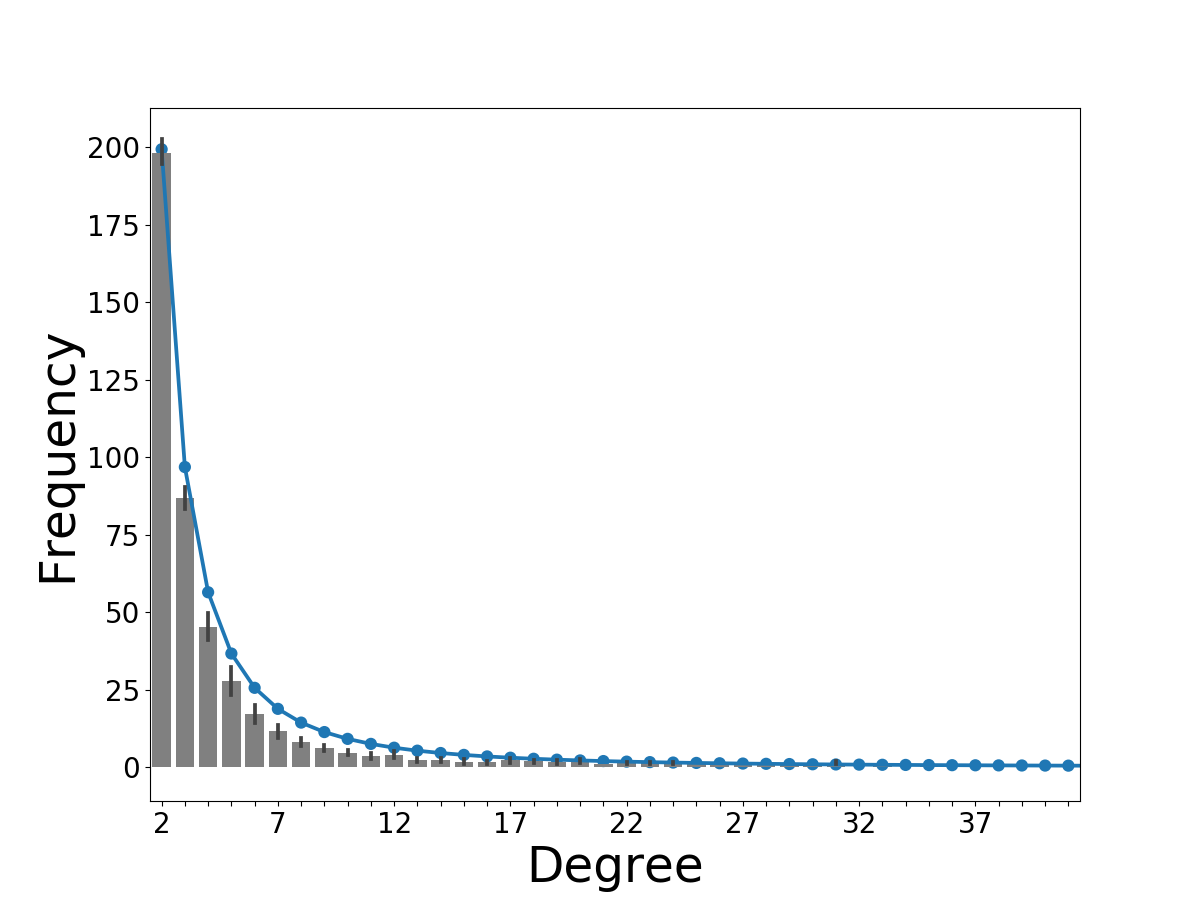}
  \caption{Connect to two nodes.}
 \label{fig:PA2}
  \end{subfigure}
  
\begin{subfigure}{0.49\textwidth}
  \includegraphics[width=\linewidth]{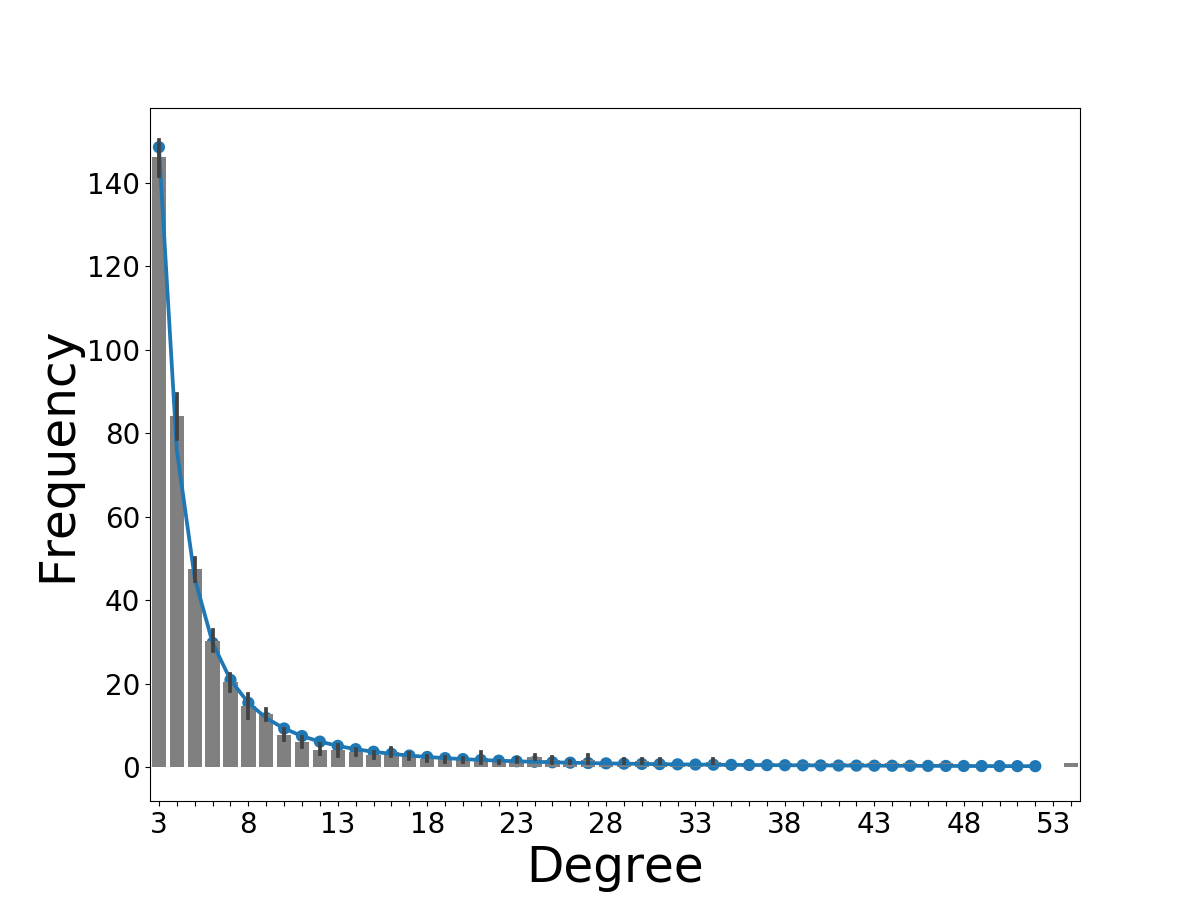}
  \caption{Connect to three nodes.}
 \label{fig:PA3}
  \end{subfigure}
\begin{subfigure}{0.49\textwidth}
  \includegraphics[width=\linewidth]{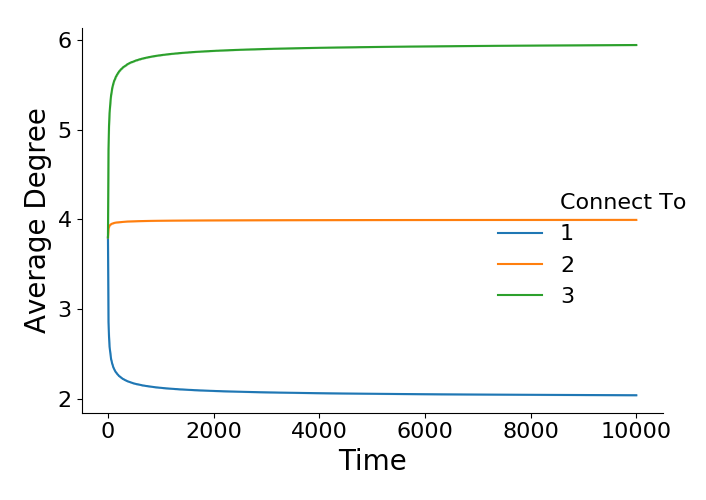}
  \caption{Average degree evolution for the three submodels.}
  \label{fig:PA4}
  \end{subfigure}

\caption{Preferential Attachment model with three connectivity strategies. }
\label{fig:PA}
\end{figure}

The experiments in \autoref{fig:PA}, were done for a virtual time of 10000 units. The generated network presents 10000 connected nodes. Three experiments were done using different values of the CONNECT\_TO parameter for 10 realisations each. Every node connects to either one, two or three existing nodes sorted from the size-biased distribution.

In \autoref{fig:PA1}, \autoref{fig:PA2} and  \autoref{fig:PA3} we show the final degree distribution of the nodes in the network. In blue we provide the curve fitting of the corresponding power-law distribution. Additionally, \autoref{fig:PA4} shows the time evolution of the network's average degree for all the experiments.

% \begin{figure}[h]\label{PA}
% \caption{Preferential Attachment degree power law}
% \centering
% \includegraphics[width=0.5\textwidth]{imgs/preferential_attachment_degree_connect_3.png}
% \end{figure}

The canonical PA model describes how some particular explicit dynamic structures evolve through time. During the implementation of this example, it was possible to analyse the impact of upward causation  in the implementation of dynamic structures based on macro-level variables. The model shows an emergent pattern that can be analysed either live (if an atomic model would require so) or, as we show in this brief example, after the simulation execution. In this sense the PA model exhibits weak emergence through the upward causation mechanism. Moreover, when designing agents that consume macro-level topology aggregated variables, it is possible to exhibit strong emergent properties.

Finally, the PA is a simple model to implement in EB-DEVS that functions as a template for the development of complex models that require a dynamic and explicit communication topology.  We will extend the SIR model in \autoref{sec:sir} to show the advantages and power of this approach.

\subsection{Model heterogeneity and strong emergence in the Sugarscape model}

In the 90s Joshua M. Epstein and Robert Axtell presented the Sugarscape model \citep{Epstein2018GrowingSocieties} to analyse wealth distribution in a grid of moving agents. As Schelling's model, agents move in a grid according to some rules. In this case, positions in the grid have a resource that agents consume (e.g.: sugar). Every time an agent moves, it looks for the position with the maximum amount of sugar within its range of vision, consumes all the sugar in that place and adds it up to its wealth. Immediately after, the agent consumes its wealth at a certain metabolic rate. If its wealth reaches zero after this consumption the agent dies. 

The age of an agent increases steadily, eventually leading to its death when a maximum value (randomly defined for each agent) is reached. When an agent dies, another agent is created at a random position (with random parameters, see below). 

Resources are distributed in the grid and regenerate at a rate of 1 unit per time unit. Each cell has a maximum capacity and sugar regenerates until the cell is full.

\subsubsection{Model dynamics}

The model has several parameters defining the agents' properties: the range of vision, metabolic rate, maximum age, and initial wealth. The agents have a state defined by their age, wealth, and their positions in the grid. The initial state is selected randomly using a uniform distribution between values defined by the model parameters. Furthermore, when an agents dies the new agent is assigned with new random values. 

The grid has another set of parameters: the number of cells, the dimensions, and the maximum capacity distribution. This parameters define the simulated terrain. In the original Sugarscape model the grid was developed to have two regions with high maximum capacity, connected by a valley of middle capacity, and surrounded by a desert of zero capacity. We use this very same terrain definition.

It is well known that the network topology influences the dynamics of the simulated model. Agents would prefer to move to positions with higher sugar availability, while positions with lower sugar availability would be statistically avoided.

For the implementation of this model we considered that the time step for resource's regeneration and consumption should be independent. This motivates the two types of atomic models present in the model, the Agents and the Cells.

\subsubsection{Emergence and adaptive behaviour in the Sugarscape model}

In the pseudocode of \autoref{lst:epstein} we present the Sugarscape model extended to include strong emergent properties. This new model makes use of a macro-level property: the Gini index, a statistical dispersion metric that is widely used to measure wealth inequality. This value is calculated at the coupled model, and its updates are triggered by the upward causation mechanism.

We extend the agent behaviour to use this information via the downward information mechanism. Each time an agent moves to a new position, and before consuming the available sugar, it queries the environment for the Gini index variable. If the Gini index is below a certain threshold, then the agent performs its regular consumption behaviour. Otherwise, if the Gini index is above the threshold (wealth is not enough equally distributed) then the agent performs a less aggressive behaviour, consuming only an amount of sugar enough to match the demand of its metabolic rate. 

The remaining features and rules of the model remains the same.

In \autoref{lst:epstein} we present the model's pseudocode.

\begin{lstlisting}[style=base,
    label={lst:epstein},
    caption={Pseudocode of the Epstein's sugarscape model. Micro-macro interactions are highlighted with red for $v_{down}$ and blue for $y_{up}$ and $x^b_{micro}$.}]
AgentState:
  # Initial agent values
  position = (randint(0,30), randint(0,30))
  vision = random(1,6)
  metabolic_rate = random(1,2)
  wealth = random(5,25)
  max_age = random(5,25)
  age = 0
  last_consumed = 0

state = AgentState()
AgentAtomicInternalTransition():
  last_position = state.position
  # Get highest sugar cell from parent
  @cell_pos, cell_sugar = $\color{red}v_{down}$(MAX_SUGAR_NEXT_CELL, state.position, state.vision)@
  state.position = cell_pos
  # Get macro-level variable, Gini index
  @gini_coef = $\color{red}v_{down}$(GINI)@
  if gini_coef < GINI_CUTOFF:
    # Regular Sugarscape behaviour
    state.last_consumed+= cell_sugar
  else:
    # If there is inequality, consume as little as metabolic_rate
    state.last_consumed+= min(cell_sugar, state.metabolic_rate)

  # Update agent state
  state.wealth+= state.last_consumed
  state.wealth-= state.metabolic_rate
  state.age+= elapsed_time
  # If the agent has no wealth or is too old, die
  if state.wealth < 0 or age > max_age:
      die()
  # Trigger the upward causation event with the agent's new state information
  /*$\color{blue}y_{up}$ = (AGENT, last_position, state.position)/*

AgentAtomicOutputFunction():
  # Consume the cell position sugar, informing the CellAtomic it was consumed
  dict_consumed_position = {state.position : self.last_consumed}
  return dict_consumed_position

AgentAtomicTimeAdvance():
  # Sample time for the next internal transition from the distribution (if the agent is alive). Otherwise it will not transition anymore.
  return random_exponential(0.5) if state.alive else INFINITY

CellState:
  # Cell initial values based on the distribution of grid's capacity values.
  position = INIT_VALUE_pos
  growth_rate = INIT_VALUE_gr
  sugar = INIT_VALUE_sg
  max_capacity = INIT_VALUE_mc

CellAtomicExternalTransition(consume):
  # Discount the consumed sugar
  state.sugar-= consume

CellAtomicInternalTransition():
  # Regenerate the sugar during the internal transition
  state.sugar+= 1
  /*$\color{blue}y_{up}$ = (CELL, state.position, state.sugar)/*

CellAtomicTimeAdvance():
  return random_exponential(0.5)

CoupledState:
  # Define the grid and the number of agents on the system
  grid = GRID(30,30)
  # Create 50 agents.
  init_agents(50)

CoupledGlobalTransition($e_g$, $\color{blue}x^b_{micro}$, $\color{red}s_{Gmacro}$):
  # Update macro-level state using agent's and cell's informed values
  for x in /*$\color{blue}x^b_{micro}$/*:
    # If the information comes from an Agent model
    if x[0] == AGENT:
      last_pos = x[1]
      actual_pos = x[2]@
      # Move the agent from the last know position to the new position if and only if the agent is alive
      $\color{red}s_G$.agent_grid[last_pos] = EMPTY
      $\color{red}s_G$.agent_grid[actual_pos] = OCCUPIED if x[3] else DEAD@
    elif x[0] == CELL:
      # If the information comes from a Cell model
      # Update the cell's available sugar
      pos = x[1]
      sugar = x[2]@
      $\color{red}s_G$.sugar_grid[pos] = sugar@

$\color{red}v_{down}$(PROPERTY, parameters):
  # Downward information function
  if PROPERTY == MAX_SUGAR_NEXT_CELL:
    # Return the cell with maximum sugar
    return get_sugar_cell(parameters.position, parameters.vision)
  if PROPERTY == GINI:
    # Return the Gini index based on the agents' wealth
    return Gini([ ag.wealth for ag in agents ])
\end{lstlisting}

\subsubsection{Experimental results and discussion} 

\begin{figure}[thb!]
\begin{centering}
    \includegraphics[width=.55\textwidth]{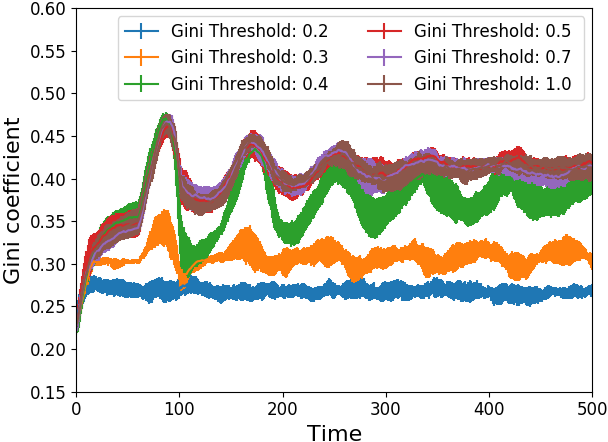}
    \vspace{1pt}
    \caption{\textbf{Simulation and modelling of the Epstein's Sugarscape model.} Scenarios varying Gini cutoff.}
    \label{fig:sugarscape}
\end{centering}
\end{figure}

As it can be seen in \autoref{fig:sugarscape}, we swept the Gini threshold to analyse the effect in the wealth distribution. The classic behaviour can be appreciated when the Gini threshold is equal to 1. In this case, the Gini index starts around 0.22, and it oscillates through time until it reaches a stable value of 0.42. The initial value depends on the first randomly assigned values of wealth. The final value depends on these values, but also is heavily dependant on the grid topology. This is also noted by Epstein and Axtell. In addition, when we run the simulations using thresholds of 0.4 or below, the Gini index oscillates and stabilises around a wealth distribution equal to the Gini threshold. Finally, if we set the threshold to 0.2 the agents behave less aggressively. In this case, the Gini index does not oscillate, reaching a final value of 0.27. Higher values of the Gini threshold don't affect the system's convergence.

In this example several complex systems' features are present. In particular, we would like to highlight how explicit and implicit communication structures can coexist.

Explicit communication channels were used between the agents and the cells. This allows the agents to signal sugar consumption, causing the cell to update its stock. This strategy facilitates the modelling by using the Separation of Concerns (SoC) principle (coined by \citet{Dijkstra1982OnThought}) which simplifies the design of complex processes by means of modularization.

Additionally, indirect communication was used to communicate to the agents the available sugar in the grid. In this regard, we observe great flexibility in two aspects. Firstly, the information on resources' availability is shared through the environment using stigmergy to coordinate the agents.
Secondly, macro-level variables such as the Gini index are integrated with ease into the model providing for a rich extension with minimal changes. For the extension is vital the role that plays upward causation for the calculation of the aggregated information and downward information by making available such information.

Finally, we remark that the model presents strong emergence, relating the micro and macro levels. The closed feedback loop uses the Gini index as a macro-level state feeding the micro level models with information that prevents the wealth accumulation. This can be observed in the model's dynamics where lower values of the Gini threshold parameter influences the average wealth distribution of the population. For higher threshold values, the system is less influenced, behaving more as the original Sugarscape model.

\subsection{A dynamic network SIR model using  a Preferential Attachment contact process}
\label{sec:sir}

SIR models are now very famous, a finite population is divided in three compartments or states: Susceptible, Infected and Recovered. Individuals in the first state may contract a disease and transition to the Infected state, where they propagate the illness to potential contacts. After an infection period, agents go to Recovered state and remain in it forever. 

We simulate an epidemic contact process taking place over a Configuration Model random graph where agents can be in three possible states: Susceptible, Infected and Recovered. The total number of Susceptible, Infected and Recovered nodes over time is denoted $S_t,I_t$ and $R_t$ respectively.

The model resorts to the so called \textit{principle of deferred decisions} \citep{Mitzenmacher2005ProbabilityAC}, revealing the
graph simultaneously with the propagation of the disease. Succinctly, the configuration model graph is dynamically generated as the infection spreads through the model's nodes. The generation of the graph follows a Preferential Attachment (PA) process with different parameters than the ones discussed in \autoref{sec:pa}.
%regarding the types of the edges
%connecting the different states of the individuals. 
% This is possible since the random environment for the epidemics' dynamics acts over a Configuration Model which is constructed using the Preferential Attachment (PA) mechanism. Nevertheless, the PA process defined for this model is subtly different as we will explain below.

\subsubsection{Configuration model graph}

The selected dynamic communication structure (stochastic environment) is the Configuration Model (CM), a class of random graph introduced in \citet{bollobas1998random}. It can be constructed as follows:

Define the sequence of degrees $s = k_1, \dots, k_n$ for nodes $i \in [1, n]$, to be independent and identically distributed according to  $p^{(n)}=(p^{(n)}_k)_{k=1,...,n}$. Following $p^{(n)}$, generate a list of half-edges for each $k_i$.  To generate the nodes degree sequence, select two random nodes with unmatched half-edges and connect them, remove those half-edges and continue selecting nodes until there are no more available nodes with unmatched half-edges.

The result of this procedure is a graph with a node degree sequence  distributed according to $p^n$, and the probability that a neighbour has degree $k$ is $\frac{k p_k}{\sum_{j=1,...,n}jp_j}$, which is the size-biased degree distribution.
This distribution will greatly influence the dynamics on CM models as opposed to the mean field point of view, where the underlying graph of potential connections is complete.

\subsubsection{Model dynamics}

We first introduce a stochastic description of the model's dynamics, later we give the ordinary differential equations of the model, and finally we describe the EB-DEVS implementation.

For a given Susceptible node we consider multiple exponential clocks with parameter $\beta$ one for each edge connecting it with an Infected node. This represents the potential encounters.
When either clock rings, the Susceptible transitions into the Infected state and remains infectious for an exponential time with mean $1/\gamma$. When this clock rings, it changes its state to Recovered  preventing future infections.

% For a given Susceptible node we consider one exponential clock with parameter $\beta$ for each edge connecting an Infected node (i.e., potential encounters). When a clock rings, the Susceptible makes a
% transition to state Infected and remains infectious during an exponential time with mean
% $1/\gamma$, whereupon it will not longer infect any longer, changing its state to Recovered.

% In \autoref{fig:SIR1} will compare the agent based simulation with the numerical integration of %the following ODE system that describe the dynamics of a SIR model over a large Configuration Model \cite{decreusefond2012large,Volz}:
% the finite-dimensional differential system that describes the evolution of the main variables that describe the epidemics, namely the
% number of individuals in each compartment ($S$, $I$, $R$), and the probability of
% interaction between the susceptible population with agents in different compartments.

Now let us introduce the differential equation model that describes the SIR  dynamics in random networks with heterogeneous connectivity \citep{decreusefond2012large,volz2008sir}. 

First consider that $p^X$ denotes the probability that an edge connects a Susceptible with a node in state $X$, for $X=S$, $I$ or $R$. Second, consider the  probability generating function  $g$ of the initial degree of agents. Finally, consider the probability $\alpha$ of a node of degree one to remain Susceptible at time t. This system strongly depends on $g$ and $\alpha$.

\begin{equation}\label{MyEq}
\left\lbrace
\begin{array}{l}
\dot{\alpha}=-\beta p^I\alpha\\
\dot{I}=-\gamma I + \beta p^I\alpha g'(\alpha)\\
\dot{p^S}=-\frac{\alpha g''(\alpha)}{g'(\alpha)}p^S\beta p^I+p^S\beta p^I\\
\dot{p^I}=-\gamma p^I+p^Irp^S\frac{\alpha g''(\alpha)}{g'(\alpha)}-rp^I(1-p^I) \\
\end{array}
\right.
\end{equation}
where $\beta$ and $\gamma$  are the contagion and recovery rates, respectively.

% The heterogeneity of the population connections is an important feature in many epidemic propagation models
% \citep{moreno2002epidemic,RevModPhys,anderson1992infectious,turnes2014epidemic}.  In our simulation we consider a random graph with degree distribution Poisson of parameter $\lambda$, arriving to a network of type Erdôs-Renyi \citep{RevModPhys}.

% In the sparse Erd\"{o}s-Renyi model, the number of neighbors in the graph follows a
% binomial distribution, which can be approximated in a large population by a Poisson
% distribution. On the other hand, when the graph is fully connected and the contact process is
% determined by a Poisson process, the number of neighbors with whom each node effectively
% connects is also Poisson distributed.

% can be grasped here through the generating function $g$ of the degree distribution, through the expression \[\frac{\alpha_tg''(\alpha_t)}{g'(\alpha_t)}\] which involves derivatives of the generation function $g$ and corresponds to the mean excess degree of the nodes, namely, the average number of outgoing links from a node reached by following a link \citep{PhysRevE.64.026118}.

The implemented EB-DEVS model translates the stochastic process in the following manner:

\begin{itemize}
    \item The agents have a state defining the compartment they belong to, the neighbours' state, and the free degree available for new connections (or available half-edges).

\item Initially, all agents are Susceptible except for a seed agent in Infected state. The Infected agent is connected to a set of randomly selected Susceptible agents.

\item Agents change their state to Infected upon reception of the 'infect' message. This is defined in the external transition function. If an agent gets Infected, it will share its new state with the connected (neighbouring) agents. When an agent changes its state to Infected, the state change triggers an upward causation event generating a structural change. 

\item The output function, which is invoked with each internal transition, emits  values to infect agents and to propagate its state when it changes. An Infected agent will either infect or recover with a probability proportional to the number of Susceptible neighbours. 

\item The atomic time advance function defines the exponential clocks for each state (Susceptible, Infected or Recovered). In the case of Infected agents, we resort to the exponential race formula \citep[lemmas 1.2.2 and 1.2.3]{Bocharov2011QueueingTheory} to trigger the time advance on the first exponential clock. The formula is adjusted for the number of Susceptible agents in that particular moment.

\item The coupled model drives the dynamic structure changes that connects the newly Infected agent to other agents using the size-biased distribution.

\end{itemize}

\subsubsection{Emergence and adaptive behaviour in the social isolation scenario}

We extend the SIR model previously discussed to highlight the strengths of the EB-DEVS approach.

To do so, we define two variables that control if agents isolate from the sources of infection or behave as previously defined. The Quarantine Threshold (QT) models when the system enters a quarantined state, and the Quarantine Acceptance (QA) models the \textit{acceptance} of the social isolation policy (enforced by the environment).

Regarding the QT parameter, if the percentage of the infected population reaches QT, the macro-level state QUARANTINE\_CONDITION is activated. If this condition is met, the agents throw a weighted coin with probability equal to the QA parameter. If the agent accepts the isolation, it prevents the infection by discarding any infect message received.

This mimics the process of social isolation during an epidemics' outbreak.

In \autoref{lst:sir} we present the pseudocode of the complete model.

\begin{lstlisting}[style=base,
  label={lst:sir},
  caption={Pseudocode of the SIR Model including the social isolation scenario. Micro-macro interactions are highlighted with red for $v_{down}$ and blue for $y_{up}$ and $x^b_{micro}$.}]
AgentState:
  id = incremental id
  # Define the initial state for each agent. There is only one infected agent at the beginning of the simulation.
  state = Susceptible if id!=0  else Infected
  neighbours_state = {}
  free_degree = random_poisson(lambda)

state = AgentState()
AgentAtomicInternalTransition():
  # State change into Recovered
  if state == Infected and to_recover:
    state = Recovered
    share_state()
    ta = INFINITY
  else:
    # Restart to_recover and infect exponential clocks
    set_infection_values()
  # Upward causation message to sync the agent state with the coupled model
  not_new_infected = False
  /*$\color{blue}y_{up}$ = (state.id, state.outdegree, not_new_infected)/*

AgentAtomicExternalTransition(message):
  share_state = False
  new_infected = False
  # If received an infect message and is Susceptible, get infected
  quarantine_accept = random() < QUARANTINE_ACCEPTANCE
  if message == 'infect':
    @do_quarantine = $\color{red}v_{down}$(QUARANTINE_CONDITION)@
    if state == Susceptible and (not do_quarantine or quarantine_accept):
      state = Infected
      share_state = True
      state.model_transition = True
      new_infected = True
  else:
    # Update the neighbours known state
    update_neighbours_state()
  # Upward causation message to sync the agent state with the coupled model
  /*$\color{blue}y_{up}$ = (state.id, state.outdegree, new_infected)/*
 
AgentAtomicTimeAdvance():
  # Time advance for the infected agent
  if state == Infected:
    if any(susceptible_neighbours):
      # If there are any susceptible neighbours
      prob = RHO_PROB / (number_of_susceptible_neighbours * BETA_PROB + Parameters.RHO_PROB)
      to_recover = random() < prob
      ta = random_exponential(1/(number_of_susceptible_neighbours * BETA_PROB + RHO_PROB))
    else:
      to_recover = True
      ta = random_exponential(1/RHO_PROB)
  # If the mode is not infected it will wait indefinitely
  elif state == Susceptible or state == Recovered:
    ta = INFINITY

AgentAtomicOutputFunction():
  ret = {}
  # If sharing its own state, share it with all the neighboring models
  if state.share:
    for outport in self.OPorts:
      ret[outport] = (state.id, state.state)
    return ret
  # If Infected and not recovering, send an infect message
  if self.state.state == SIRStates.I and self.state.to_recover == False:
    # Ignore the case where there are no neighboring agents
    if len(self.OPorts) == 0:
      return 
    susceptible_ids = state.susceptible_neighbors_ids
    if susceptible_ids == None:
      return 
    # Randomly choose one susceptible neighbor
    outmodel_id = random_choice(susceptible_ids)
    outport = get_outport(from=state.id, to=outmodel_id)
    if outport != None:
      ret = {outport[0]: "infect"}
    return ret
  return 

CoupledState:
  topology = Parameters.INITIAL_TOPOLOGY
  nodes_degree = {}
  agent_states = {}

$\color{red} s_G$ = CoupledState()
CoupledGlobalTransition($e_g$, $\color{blue}x^b_{micro}$, $\color{red}s_{Gmacro}$):
  # Update the out degree for the node with id=id
  for id, outdegree, new_infected in /*$\color{blue}x^b_{micro}$/*:
    nodes_degree[id] = outdegree
    # If a new infected node, then change the structure
    if new_infected:
      # Structure change code section
      # Generate the probability distribution for nodes connection
      degree_sequence = nodes_degree.values()
      pk = [degree / sum(degree_sequence) for degree in degree_sequence]
      # Pick one with the discrete distribution
      connect_to_node = choice(nodes_degree.keys(), pk, replace=False, number=Parameters.CONNECT_TO)
      # Connect the newly_infected to each of the selected nodes (all susceptible)
      for node in connect_to_node:
        connect_nodes(id, node)
      update_topology()

$\color{red}v_{down}$(PROPERTY, parameters):
  if PROPERTY == ENVProps.QUARANTINE_CONDITION:
    # count the number of agents per state.
    (unique, counts) = unique_count($s_G$.agent_states) 
    # Filter the infected agents.
    unique_counts_dict = dict(zip(unique, counts))
    infected_number = unique_counts_dict['Infected']
    # Calculate the percentage of infected agents.
    infected_percentage = infected_number / float(len(self.agents))
    # Return true if the agents should isolate.
    return QUARANTINE_THRESHOLD < infected_percentage
      
\end{lstlisting}

\subsubsection{Experimental results and discussion} 

\begin{figure}[htb!]
    \centering % <-- added
  \includegraphics[width=0.8\linewidth]{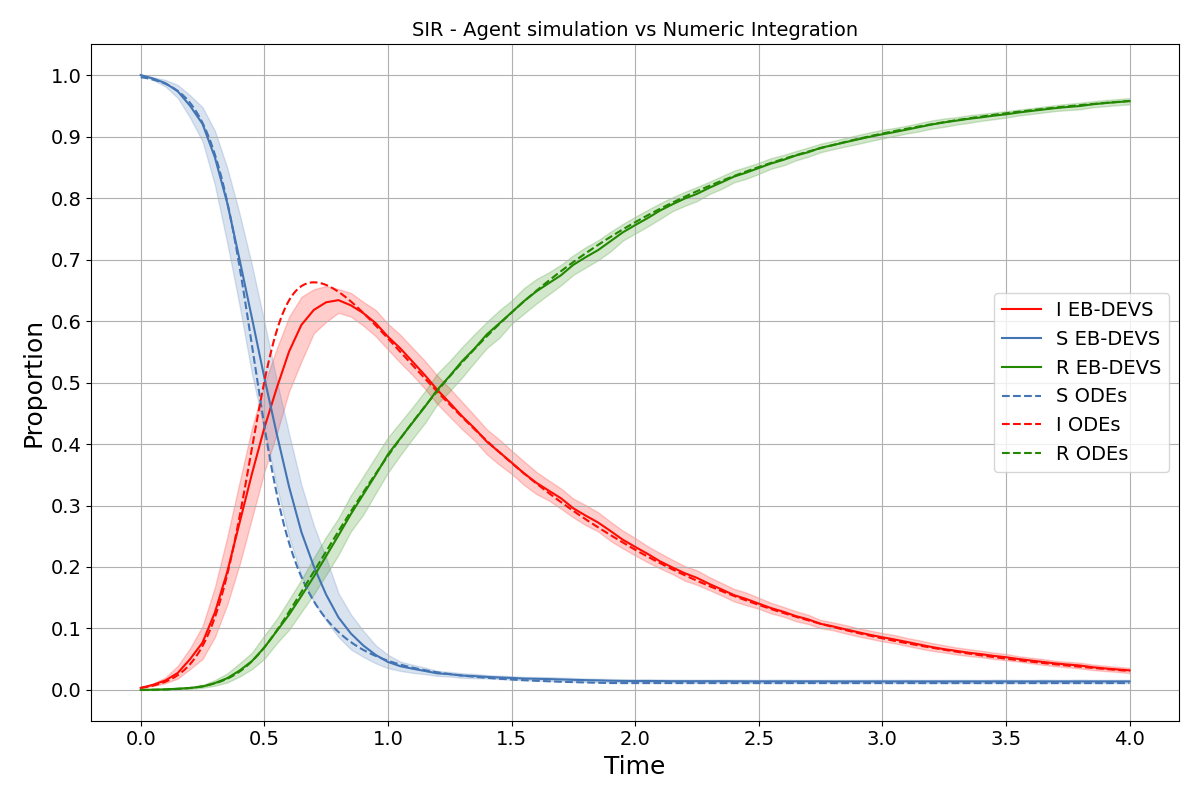}
\caption{Comparison of SIR dynamics for EB-DEVS simulation (solid line) vs. numerical integration of ODEs (dashed). Parameters: $\gamma=1$, $\beta=3$, $\lambda=8$}
  \label{fig:SIR1}
\end{figure}

In \autoref{fig:SIR1} we show how the agent-oriented EB-DEVS model compares to a mean field model (a set of ODEs solved by numerical integration) for the same SIR system.

The experiments were configured for parameters $\gamma=1$, $\beta=3$ and an initial free degree distribution of nodes following a $Poisson(\lambda=8)$. The experiments were run for 4 units of time and 30 realisations.

We can observe that the EB-DEVS model matches closely the theoretical trajectories of SIR dynamics over a Configuration Model \citep{decreusefond2012large}.

\begin{figure}[htb!]
    \centering % <-- added
\begin{subfigure}{0.45\textwidth}
  \includegraphics[width=\linewidth]{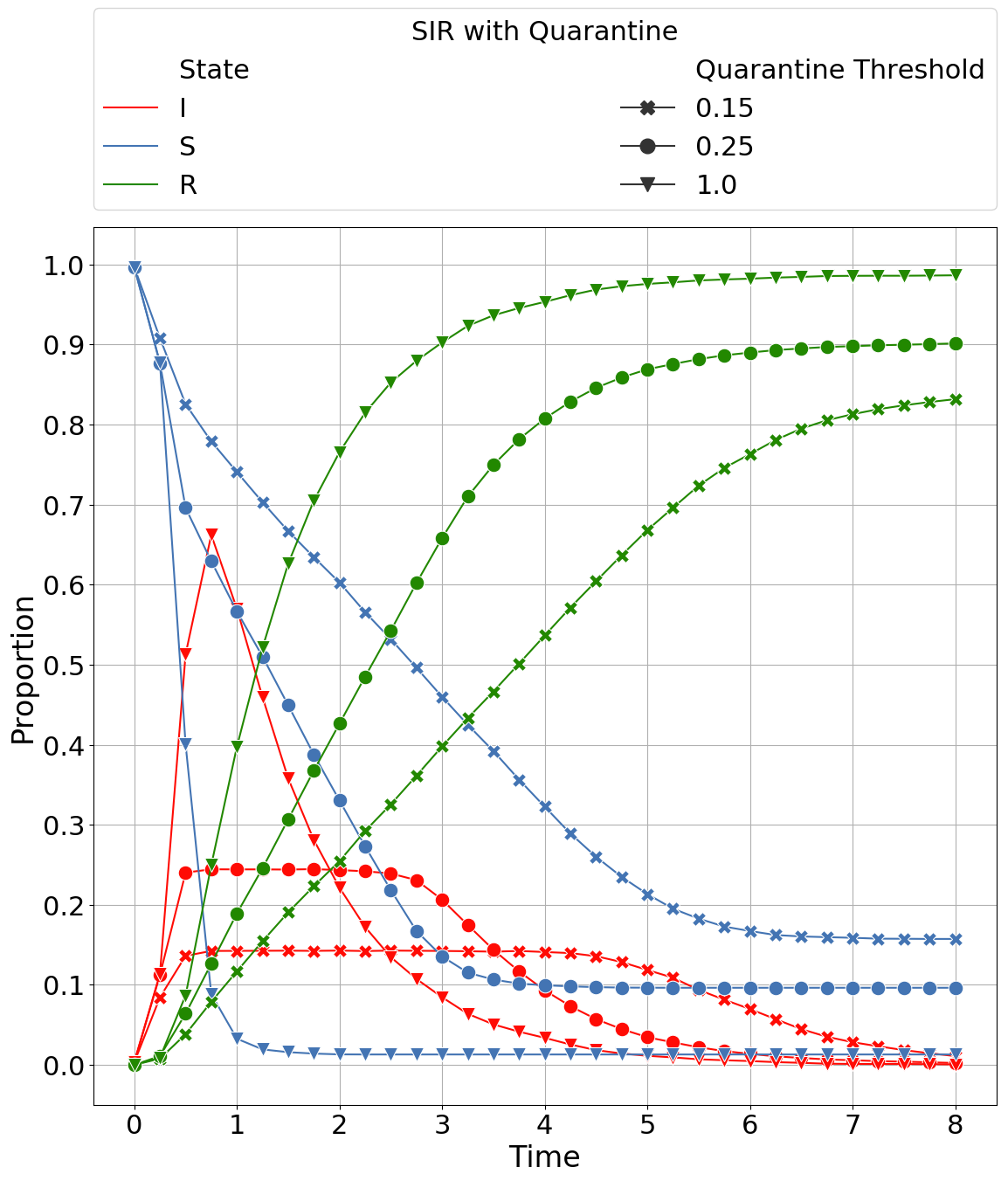}
  \caption{Quarantine Threshold (QT) sweep. $QA=1, \gamma=1$, $\beta=3, \lambda=8$}
  \label{fig:SIRQT}
\end{subfigure}\hfil % <-- added
\begin{subfigure}{0.45\textwidth}
  \includegraphics[width=\linewidth]{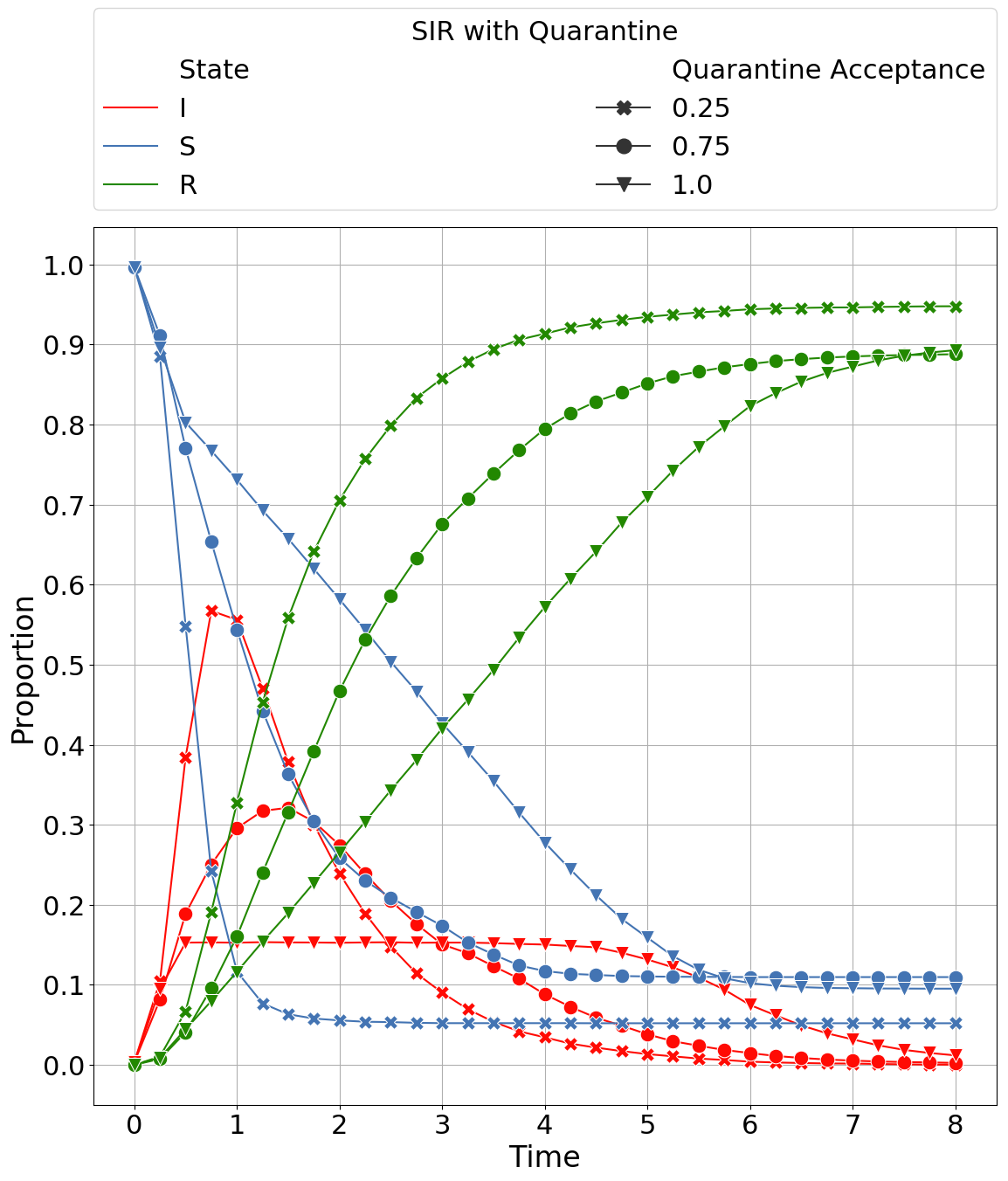}
  \caption{Quarantine Acceptance (QA) sweep. $QT=0.15, \gamma=1, \beta=3, \lambda=8$}
  \label{fig:SIRQA}
\end{subfigure}
\caption{Emergent behaviour for a SIR model with quarantines.} 
\label{fig:SIREXP2}
\end{figure}

Regarding the \textit{social isolation} model, we can see from the scenarios evaluated in  \autoref{fig:SIRQT} that lower QT values correlate with a reduction in the maximum infected population.

By preventing social contact during the rise of an epidemic outbreak it is theoretically possible to control the simultaneous number of infected agents. This number will depend on the chosen parameters QT and QA. Nevertheless, this strategy shows that in time all agents get infected.

% This strategy could prevent the health system to be overloaded with infected patients. %, thus enabling Intensive care units (ICUs) to work at full capacity without running out of resources.
% This can be done establishing Quarantines depending on the observed percentage of  infected people. %Quarantine Threshold in this case.

We observe that the  QA parameter plays a fundamental role in the evolution of the  infected population. As it can be seen in \autoref{fig:SIRQA} the maximum number of infected agents depends not only on the QT parameter, but is also conditioned by QA. Lower  QA values match the standard SIR model behaviour, while a higher QA present a behaviour similar to the exhibited in \autoref{fig:SIRQT}.

We also observe a plateau in the curve of the infected population in \autoref{fig:SIRQT} which is not characteristic of classical SIR model. Indeed, quarantines can control the peak of epidemics, demonstrating that such strategies can be useful to avoid saturations of the health system.

In this example multiple complex system's features converge. We shall highlight how the combination of such features expand the modelling capabilities.

The utilisation of explicit and dynamic structures, together with the modelling of weak emergence, allowed for the description of complex infection dynamics over evolving networks. This allows for modelling the spread of an infection as a contact process using explicit links that causally affect models (agents) at a same spatio-temporal level. This can be seen in the SIR-CM model, where the contact network is revealed as the system evolves. The utilisation of the upward causation mechanism is key for the development of such network: the size-biased distribution is generated based on the observed micro-level information.

In this experiment we stepped away from analysing the network structure impact over the epidemics.  This analysis can be found in \citet{PhysRevE.64.026118,RevModPhys,ferreyra2019sir}.

In regard of the social isolation model, the downward information mechanism closes the micro-macro level loop presenting strong emergence properties that are observed in \autoref{fig:SIREXP2}. These macro-level structures are a consequence of the micro-macro level interactions.

\section{Discussion}
\label{sec:discussion}

By visiting different types of models for varied social systems we saw how archetypal structural features (see \autoref{tbl:models_features}) can be modelled and implemented with EB-DEVS, and their relation with emergent properties.

In the Dissemination of Culture model, by means of the \textit{Fads and Fashion} extension, we showed that introducing emergent properties (which impact noticeably the model's phase transitions) can be done in a compact and straightforward way, with minimal additions to a reference baseline model. In the example, we add strong emergence by including a simple rule that modifies the agents' adaptive dynamics.

Afterwards, in the Segregation model we used an implicit communication structure, relying on the upward causation mechanism, to implement the agents' coordination. This pattern allowed us to introduce weak emergence and stigmergy in the model. 

Later on, in the Preferential Attachment and SIR models, we showed how evolving networks can produce complex behaviours. These behaviours are easily expressed with simple rules in EB-DEVS based on the state of macro-level variables.

Finally, with the Sugarscape model we showed how different types of models at the microscopic level can be integrated using implicit communication mechanisms. This is, we established a well-defined separation of concerns between atomic models of type \textit{Cell} and \textit{Agent}, and defined their rules of interaction by means of algorithms at the global transition function. Furthermore, the addition of strong emergent behaviour allowed for the definition of macro-level variables that influence back the behaviour of micro-level models.

Considering the implemented models, we highlight the following salient aspects:
 
\begin{itemize}
  \item EB-DEVS helps the modeller to deal with systems requiring a multilevel treatment by relying on a well-defined multilevel framework featuring unambiguous simulation rules. This avoids the development of ad hoc single-use mechanisms to resolve the interaction between the micro and macroscopic levels.
  \item The decision to resort to closed micro-macro feedback loops or to use only partial micro-macro communication (upstream or downstream) can be made at any point in the modelling process with minimal impact on the previously modelled dynamics.    
  \item Evolving from models without emergent behaviour to extensions that include some form of emergence is a straightforward task that preserves the structure of the base model, thus facilitating also the assessment of the impact of emergence processes compared to more simplified versions of a same system model.
  \item Implicit and explicit communication structures can be easily integrated together to design elaborated coordination mechanisms. These communication patterns are often very useful while designing complex systems with adaptive agents.
  \item Dynamic communication structures (i.e., those that evolve with time) span evolving networks that are often seen in natural, social and engineered systems. In ABMs these structures require a higher level entity to manage the connections. This can be naturally modelled with EB-DEVS via the macro-level state.
\end{itemize}

\section{Conclusions}
\label{sec:conclusions}

In this paper we presented several classical models in computational sociology and implemented them with the EB-DEVS formalism. The chosen models are often considered as typical examples of complex systems and can exhibit various forms of emergent behaviour.

Throughout these case studies we have identified certain structural features by which the systems under study can be typified, and we show the relationship of these features to the activity of modelling emergent behaviour with EB-DEVS.

We showed different modelling practices that facilitated the modelling of explicit and implicit communication structures, static or dynamic, with or without micro-macro interaction, and with weak or strong emergent behaviour (in the latter case, accepting the live identification of the emergent property). In each example we discussed the role of communication structures in the development of multilevel simulation models, and illustrated how micro-macro feedback loops enable the modelling of macro-level properties. 

Complex systems modelling is a necessary tool to explore and understand social processes by means of computer simulation, and it benefits from having formal mechanisms that allow global-level properties derived from local-level interactions. In this paper we showed how EB-DEVS permits conceptualising the analysed societies by incorporating emergent behaviour when required, for example by incorporating the concept of Gini in the Sugarscape model, Fads and Fashion in the Dissemination of Culture model or Quarantines in a SIR epidemic model.

However, although EB-DEVS has proven to be a convenient formalism for a wide variety of complex systems, further experimentation is needed. It remains necessary to demonstrate that the approach works as expected in multi-layered systems where micro-macro information traverses the system hierarchy across more than 2 levels (a feature common to all models exercised in this work). In addition, we have not yet explored the impact on simulation performance, which can be a key decision factor when selecting a M\&S technology (especially for complex systems that are often composed of a large number of elements at the microscopic level).

\bibliographystyle{unsrtnat}
\bibliography{biblio.bib ,  references_mendeley_priv} % Please set the right name for your bib file

%%%%%%%%%%%%%%%%%%%%%%%%%%%%%%%%%%%%%%%%%%%%%%

\end{document}